\documentclass[useAMS,usenatbib,usegraphicx]{mn2e}

\def\gtsim {>\kern-1.2em\lower1.1ex\hbox{$\sim$}~}   
\def\ltsim {<\kern-1.2em\lower1.1ex\hbox{$\sim$}~}   

\def \apj {ApJ}
\def \apjs {ApJS}
\def \aj  {AJ}
\def \aap {A\&A} 
\def \mnras {MNRAS}
\def \pasj {PASJ}
\def \araa {ARA\&A}

\title[GRAPE-SPH Chemodynamical Simulation of Elliptical Galaxies I]{GRAPE-SPH Chemodynamical Simulation of Elliptical Galaxies I: Evolution of Metallicity Gradients}
\author[Chiaki Kobayashi]{Chiaki Kobayashi$^{1}$\thanks{E-mail:
chiaki@MPA-Garching.MPG.DE}\\
$^{1}$Max-Planck-Institute for Astrophysics,
Karl-Schwarzschild-Str. 1, D-85741 Garching, Germany}

\begin{document}

\date{Accepted 3rd Oct 2003; Received 3rd Mar 2003.}

\pagerange{\pageref{firstpage}--\pageref{lastpage}} \pubyear{2003}

\maketitle

\label{firstpage}

\begin{abstract}
We simulate the formation and chemodynamical evolution of 124 elliptical galaxies by using 
a GRAPE-SPH code that includes various physical processes associated with the formation of stellar 
systems: radiative cooling, star formation, feedback from Type II and Ia supernovae 
and stellar winds, and chemical enrichment.
In our CDM-based scenario, galaxies form through the successive merging of sub-galaxies with various masses.
Their merging histories vary between a major merger at one extreme, and a 
monolithic collapse of a slow-rotating gas cloud at the other extreme.
We examine the physical conditions during 151 merging events that occur in our simulation.
The basic processes driving the evolution of the metallicity gradients are as follows:
i) destruction by mergers to an extent dependent on the progenitor mass ratio.
ii) regeneration when strong central star formation is induced at a rate dependent on the gas mass of the secondary.
iii) slow evolution as star formation is induced in the outer 
regions through late gas accretion.
We succeed in reproducing the observed variety of the radial metallicity gradients.
The average metallicity gradient $\Delta\log Z/\Delta\log r \simeq -0.3$
with dispersion of $\pm 0.2$ and no correlation between gradient and galaxy mass are consistent with observations of Mg$_2$ gradients.
The variety of the gradients stems from the difference in the merging histories.
Galaxies that form monolithically have steeper gradients, while galaxies that 
undergo major mergers have shallower gradients.
Thus merging histories can, in principle, be inferred from the observed metallicity 
gradients of present-day galaxies.
The observed variation in the metallicity gradients cannot be explained by either 
{\it monolithic collapse} or by {\it major merger} alone.
Rather it requires a model in which
both formation processes arise, such as the present CDM scheme.
\end{abstract}


\section{INTRODUCTION}
\label{sec:intro}

The internal structure of galaxies,  
spectrophotometric, chemical, and dynamical properties of 
various locations within a galaxy,  is closely related 
the processes of galaxy formation and evolution.
Stars in a galaxy are fossils;
the star formation and chemical enrichment 
history of the galaxy are imprinted on their kinematics and chemical abundances.
The SAURON project with William Herschel Telescope (\citealt{bac01}) is providing the wide-field 
mapping of the kinematics and stellar population, which will certainly give stringent 
constraints on the galaxy formation and evolution.
Multiobject and Integral Field Spectrographs are being developed also on 8-10m ground-based telescopes, which can give the time evolution of such internal structure.
To derive the physical processes from such observational data,
it is necessary to construct a realistic model, i.e., a three-dimensional chemodynamical model, 
and to compare the theoretical predictions with such observational data.

Theoretical approaches have been performed in many ways:
1) The one-zone model (e.g., \citealt{tin80}; \citealt{ari87}; \citealt{mat96}) played an important 
role in constructing the basic evolutionary scenarios of galaxies.
In this model,
the star formation history of a galaxy is constructed by carefully comparing the model predictions 
with the mean photometric and chemical properties of observed galaxies.
However, because of the simplified assumption that all matter in a galaxy is well-mixed 
instantaneously, the one-zone model provides no idea of the internal structure of galaxies.
2) The semi-analytic model (e.g., \citealt{kau93}; \citealt{col94}), which is based on the Press-Schechter 
theory, can provide the mass function of dark halos, their survival timescales, and the merging rates. 
By adopting empirical laws to determine the stellar mass and the mass of heavy elements, and by 
introducing simple rules to determine the galaxy morphology (i.e., an elliptical galaxy forms 
from the major merger of spiral galaxies), the semi-analytic model could reproduce some correlations among global properties (e.g., the Tully-Fisher relation 
and color-magnitude relation) and some constraints on the number of galaxies (e.g., the luminosity function and number counts).
However, there seems to be some difficulties to explain the number evolution of elliptical galaxies (e.g., \citealt{ben02}).
The observed information on the internal structures remains untouched.
3) Numerical simulations of dissipationless systems (e.g., \citealt{too72}; \citealt{whi78}) provide 
the interpretation of the interaction of galaxies and the internal structure of a galaxy.
The gas dynamics in three dimensions was included in such numerical 
simulations (e.g., \citealt{her89}), and then star formation, feedback (\citealt{kat92}), and chemical enrichment (\citealt{ste94}) were included.
It is now possible to predict the spatial distributions of gas, stars, and heavy elements in a galaxy.
However, the comparison with the observations has not been fully attempted yet.
The reason is that it takes long time to calculate the evolution of one galaxy with enough 
resolution to predict such distributions.
To reduce the calculation time and to improve the resolution, a variety of methods have been invented.

The Smoothed Particle Hydrodynamics method (SPH) is widely used
to calculate three dimensional hydrodynamics with a Lagrangian scheme
(\citealt{luc77}; \citealt{gin77}; \citealt{mon92} for a review), 
and has been applied to many astrophysical problems
that have large density contrasts;
formation of galaxies (\citealt{her89}; \citealt{nav93}; \citealt{ste94})
and a cloud-cloud collision (\citealt{lat85}; \citealt{hab92}).
Various codes have been developed to combine SPH with collisionless particles
(i.e., dark matter and star particles; hereafter an N-body system)
using different methods to calculate gravitational forces;
direct summations,
Particle-Particle/Particle-Mesh methods (\citealt{evr88}), 
Tree methods (\citealt{her89}; \citealt{ben90}),
and the method using the special purpose computer GRAPE
(\citealt{ume93}; \citealt{ste96}).
GRAPE (GRAvity PipE) is a special purpose computer for efficiently
calculating gravitational force and potential (\citealt{sug90}).
The GRAPE-SPH enables us to simulate the formation and evolution of a galaxy with more than 
$10^4$ particles in calculation time as short as a few days.
It makes it possible to simulate many types of galaxies with different initial conditions, 
which is crucial to study the formation and evolution of galaxies statistically.

The aim of our study is to put constraints on the formation history of elliptical galaxies by comparing the observed internal structures of stellar population.
To construct a self-consistent three-dimensional chemodynamical model, we have introduced 
various physical processes associated with the formation of stellar systems 
such as radiative cooling, star formation, feedback of Type II and Ia supernovae (SNe II 
and SNe Ia), and stellar winds (SWs), and chemical enrichment.
The chemical enrichment of SNe Ia has been recently included in several chemodynamical models.
Among two alternative scenarios of the SN Ia progenitor (e.g., \citealt{kob98}),
most chemodynamical models (\citealt{rai96}; \citealt{car98}) adopted \citet{gre83}'s formulation based on the double-degenerate scenario.
\citet{mos01} adopted single time delay for the SN Ia contribution, and the parameter corresponds to the double-degenerate scenario.
The single-degenerate scenario has been introduced in \citet{nak03} and \citet{kaw03}.
It may be useful to note that \citet{woo95}'s iron yield is too large compared with the observed abundance ratios in the Milky Way Galaxy, and thus their iron yield should be reduced to be half.
Such modification is always adopted in the one-zone models, but is never mentioned in many chemodynamical models.
We have constructed a realistic model of chemical enrichment, excluding the instantaneous recycling approximation, including the mass-dependent yields of SNe II and the single-degenerate scenario of SNe Ia.
We then solved the evolution of slowly-rotating systems that consist of dark matter, gas, 
and stars from various initial conditions to predict the spatial distribution of stellar 
population within a galaxy. By comparing the theoretical metallicity gradients 
with the observed ones, we discuss the origin of elliptical galaxies.

How elliptical galaxies form is a long-standing issue as a matter of big debate.
The regularity in the light distribution and the global velocity anisotropy in elliptical 
galaxies were explained by the violent relaxation (\citealt{lyn67}). 
Effectively dissipationless formation of an elliptical galaxy was discussed in various ways; e.g., 
stars have formed prior to the beginning of the collapse of the gas cloud
(\citealt{got73}, 1975) or stars have formed slowly in disk galaxies which 
subsequently merge to make a spheroidal galaxy (\citealt{too77}; \citealt{mar77}; \citealt{bar88}). 
The limit of completely dissipationless collapse is amenable to N-body experiments, 
which can form the objects that have the observed dynamical properties.
However, the dissipation during the formation is indispensable to explain the photometric 
and chemical properties of elliptical galaxies such as the color-magnitude 
relation, the mass-metallicity relation, and the radial metallicity gradients.

Two competing scenarios of the formation of elliptical galaxies have so far been proposed: 
Elliptical galaxies should form monolithically by gravitational collapse
of gas cloud with considerable energy dissipation 
(hereafter referred to as the monolithic collapse hypothesis; e.g., \citealt{lar74b}; 
\citealt{ari87}), or alternatively 
ellipticals should form via mergers of gaseous disk galaxies or of many dwarf galaxies 
(hereafter referred to as the merger hypothesis; e.g., \citealt{too77}; Kauffmann et al. 1993; \citealt{bau96}; \citealt{ste02}). 
The merger hypothesis can be supported by the dynamical disturbances of observed 
ellipticals such as shells/ripples and multiple cores (\citealt{sch90}; \citealt{sch92}; see also \citealt{ben92}), 
and may easily explain the morphology-density relation of galaxies in clusters 
(\citealt{dre80}; \citealt{dre97}).
However, elliptical galaxies show apparently little evidence for on-going star
formation, the bulk of their stars are old (e.g., \citealt{kod97}; 
\citealt{sta98}; \citealt{kod98}; \citealt{sil98}).
The monolithic collapse hypothesis assumes that the bulk of stars in
ellipticals form during an initial star burst at high redshift, 
and that the star formation is terminated by a supernovae-driven galactic wind 
that expels the left-over interstellar gas from galaxies. 
The galactic wind is supposed to play an essential role 
in injecting heavy elements into the hot intracluster gas (\citealt{cio91}),
and predicts tight correlations among global properties of
galaxies such as the color-magnitude relation (\citealt{bow92}),
the metallicity-velocity dispersion relation (\citealt{dav87}),
and the fundamental plane (\citealt{djo87}; \citealt{dre87}). 
(The color-magnitude relation could be reproduced also under the merger hypothesis (\citealt{kau98}, but see \citealt{col00}).)
Recent observations of clusters at high redshifts reveal that 
these relationships exist even at $z \sim 1$ 
(\citealt{dic96}; \citealt{sch97}; \citealt{kel97}; 
Stanford et al. 1998),
which indicates 
that the bulk of stars in cluster ellipticals forms at the redshift 
$z_{\rm f} \gtsim 2.5-4$ (\citealt{kod98}).

However, for cluster early-type galaxies, it has been argued that the ``progenitor bias'' is significant; the progenitors of the youngest low-redshift early-type galaxies drop out of the sample at high redshift (\citealt{kau96}; \citealt{van96}, 2001). 
Thus the evolution of field early-type galaxies is now paid attention as the observational constraints.
\citet{fra98} found that HDF-N early-type galaxies are relatively young with the formation epochs spanning $1 \ltsim z \ltsim 4$,
but no evolution of field elliptical galaxies found at least $z \ltsim 1$ (\citealt{sch99}; \citealt{bri00}; \citealt{im02}; see also \citealt{dad00}).
\citet{dro01} found mass evolution at $0.4 < z < 1.2$.
At high redshift, the number of field red ellipticals is smaller than expected by the monolithic collapse hypothesis (\citealt{zep97}; \citealt{men99}; \citealt{bar99}), and there may be some global evolution that is consistent with the hierarchical clustering scenario (\citealt{fon99}; \citealt{dic03}; but see \citealt{cim02}).

The radial metallicity gradient gives one of the most stringent constraints on the galaxy formation.
Numerical simulations of the collapse of galaxies including star formation definitely 
predict strong radial gradients in chemical enrichment (e.g., \citealt{lar74a}, 1975; \citealt{car84}), 
whereas the dissipationless collapse models predict no gradient in chemical enrichment (\citealt{got73}, 1975).
During the collapse, gas is chemically enriched, flows inward, and forms new stars, which form the radial metallicity gradients.
The metallicity gradients are observed as radial gradients of colors and spectral line 
indices (e.g., \citealt{fab73}; \citealt{fab77}; \citealt{dav93}; \citealt{kob99} for a review).
A typical observed metallicity gradient of elliptical galaxies is 
$\Delta \log Z / \Delta \log r \simeq -0.3$,
which is less steep than those predicted by
numerical simulations of dissipative collapse ($-0.35$ in \citealt{lar74a}; $-1.0$ 
in \citealt{lar75}; $-0.5$ in \citealt{car84}).
Furthermore, 
if elliptical galaxies form monolithically from a massive gas cloud,
metallicity gradient should correlate with global properties of galaxies
in the sense that more massive galaxies have steeper gradients (\citealt{car84}).

The observational feature of metallicity gradients is complicated and 
confusing because of a lack of suitable sample of uniform quality.
It was shown that 
elliptical galaxies with larger values of the central Mg$_2$ ($\sim5100\AA$) index 
tend to have steeper Mg$_2$ gradients (\citealt{gor90}; \citealt{car93}; 
\citealt{gon96}).
However, Davies et al. (1993) did not 
find any significant correlation between the Mg$_2$ gradient and $\sigma_0$
in the sample of 13 galaxies.
Kobayashi \& Arimoto (1999) re-studied line-strength gradients of 80 elliptical galaxies 
by using the indices of Mg$_2$, Mg$_{\rm b}$ (5177$\AA$), Fe$_1$ (5270$\AA$), Fe$_2$ (5335$\AA$) and H${\beta}$ (4861$\AA$), 
and found that 
the metallicity gradients do not correlate with any physical properties of
galaxies, including central and mean metallicities, 
central velocity dispersions, absolute B-magnitudes, 
absolute effective radii, and dynamical masses of galaxies.
Elliptical galaxies have different metallicity gradients, 
even if they have nearly identical
properties such as masses, luminosities, and metallicities. 

This discrepancy could be solved if mergers flatten
the original gradient. Indeed numerical 
simulations showed that the gradient in a disk galaxy should be halved after 
three successive mergers of galaxies with similar size (\citealt{whi80}).
However, according to the dissipationless N-body experiment, 
the initial state is not fully wiped out during the violent relaxation phase,
and N-body particles that were in the outer region of a progenitor galaxy are found in the similar location after merging events (\citealt{vanalb82}).
From this point, it has been mentioned that metallicity gradients is not reduced by a merger (e.g., \citealt{bar96}).
Simulations of both dissipative collapse and mergers
leave room for improvements, 
because essential physical processes such as 
star formation, feedback of supernovae, and metal
enrichment were not taken into account.

Here we simulate the chemodynamical evolution of elliptical galaxies based on the CDM picture.
In the CDM cosmology, the amplitude of primordial fluctuation decreases with increasing 
wavelength, and the formation of structure is driven by the hierarchical clustering.
Galaxies should form through the successive mergings of sub-galaxies with various masses.
Contrary to the semi-analytic models, we exclude the assumption that elliptical galaxies form only 
from the major merger of disk galaxies.
Instead we allow various merging histories for elliptical galaxies. In some cases, an 
elliptical galaxy forms by an assembly of gas rich small galaxies, which looks like a monolithic collapse.
In other cases, the evolved galaxies with little gas merge to form an elliptical galaxy.
This scenario is the midway of monolithic collapse and major merger of disk galaxies. 

In this paper, we first describe the GRAPE-SPH code and the modeling of physical processes (\S \ref{sec:model}).
We classify simulated galaxies according to their merging histories (\S \ref{sec:class}) 
and derive the present metallicity gradients (\S \ref{sec:fit}). In \S \ref{sec:grad}, 
we show that the scatter of the metallicity gradients comes from the difference in merging 
histories, and discuss the origin of elliptical galaxies by comparing with the observation.
In \S \ref{sec:evozg}, we examine the evolution of metallicity gradients via merging events 
that occur in our simulation, and manifest the dependences on mass ratios of merging galaxies, 
gas fractions, and induced star formation.
In \S \ref{sec:discussion}, we mention some future works and possible problems.
Our conclusions are given in \S \ref{sec:conclusion}.

\section{CHEMODYNAMICAL MODEL}
\label{sec:model}

\subsection{Hydrodynamics}

Here we describe the GRAPE-SPH code, which is originally written by N. Nakasato (2000), 
and is highly adaptive in space and time by means of individual smoothing lengths and individual 
timesteps. The SPH formulation used in the code is almost the same as \citet{nav93}.

Instead of the continuity equation,
the density $\rho$ of each particle is determined from
\begin{equation}\label{eq:hydro1}
\left< \rho(\mbox{\boldmath$r$}_i) \right> \equiv
\rho_i = \sum m_j W(r,h)  ,
\end{equation}
where $r$ is defined as $r \equiv |\mbox{\boldmath$r$}_i-\mbox{\boldmath$r$}_j|$, and $h$ 
denotes the symmetrized smoothing length defined as $h \equiv h_{ij}=\frac{1}{2}(h_i+h_j)$.
The momentum equation is represented as 
\begin{equation}\label{eq:hydro2}
\frac{D\mbox{\boldmath$v$}_i}{Dt} = -\sum m_j 
\left( \frac{P_i}{\rho_i^2}+\frac{P_j}{\rho_j^2}+\Pi_{ij} \right)
\nabla_i W(r,h)
-(\nabla\Phi)_i   .
\end{equation}
The energy equation is represented as
\begin{equation}\label{eq:hydro3}
\frac{Du_i}{Dt} \!=\! -\!\sum\! m_j \!\!
\left( \frac{P_i}{\rho_i^2}\!+\!\frac{1}{2}\Pi_{ij} \! \right) \!\!
\left( \mbox{\boldmath$v$}_i\!-\!\mbox{\boldmath$v$}_j \right)\cdot
\nabla_i W(r,\!h)
+\frac{{\cal{H}}_i\!-\!\Lambda_i}{\rho_i}   ,
\end{equation}
where thermal conduction is neglected.
We also use the equation of state for an ideal gas with $\gamma=5/3$;
\begin{equation}\label{eq:hydro4}
P_i=(\gamma-1)\rho_i u_i  .
\end{equation}

The smoothing length $h$ varies spatially, evolves with time, and is computed for 
each particle in every timestep.
For the kernel $W$, a spherically symmetric spline (\citealt{mon85}; \citealt{mon92}) is adopted.
For the derivative of the kernel $\nabla W$, the revised form by \citet{tho92} is adopted.
For the artificial viscous term $\Pi_{ij}$, we use the modified \citet{mon83} tensor and 
the shear free viscosity formulation with $\alpha=1.0$ and $\beta=2.0$ (\citealt{bal95}; \citealt{nav97}). 
The heating $\cal{H}$ and cooling $\Lambda$ rates are described in the next session.

In the SPH code, a gas particle interacts with dark matter and star particles only by gravity. 
The gravitational potential $\Phi$ is given as
\begin{equation}
\label{eq:epsilon}
\Phi(\mbox{\boldmath$r$}_i) = -G \sum \frac{m_j}{\sqrt{|\mbox{\boldmath$r$}_i-\mbox{\boldmath$r$}_j|^2+\epsilon^2}}  ,
\end{equation}
where $G$ is the gravitational constant. $\epsilon$ is the gravitational softening length, 
which we set $\epsilon=0.5$ and $1.0$ kpc for the high- and low- resolution, respectively.
The gravity between these particles is calculated in direct summation
using the GRAPE5 MUV (Mitaka Underground Vineyard) system in the National Astronomical Observatory 
of Japan and GRAPE6 of the University of Tokyo.

We use an individual timestep scheme with the following two steps.
First, we compute the timestep of each particle $\Delta t_{\rm e}$ from the dynamical 
criteria using the velocity $v$ and the acceleration $a$ (\citealt{kat96}); 
$\Delta t_{\rm e}$ is the minimum of $\frac{\eta^2\epsilon}{|v|}$ and $\eta\sqrt{\frac{\epsilon}{|a|}}$, where $\eta$ is 
a numerical parameter, and is set to be $\eta=0.5$.
Then, we derive the actual timestep $\Delta t_{\rm a}$ from the greatest power of 2 
subdivision of the system timestep $\Delta t_{\rm sys}$, which is smaller than 
$\Delta t_{\rm e}$; $\Delta t_{\rm a}=\Delta t_{\rm sys}/2^n \le \Delta t_{\rm e}$ (\citealt{nav93}).
The system timestep is a fundamental timestep used to synchronize all particles, and is 
set to be $\Delta t_{\rm sys}=2$ Myr.
The time integration of the equation of motion is done using a leap-frog method modified for the individual timestep scheme.

\subsection{Physical Processes}

To simulate the formation and evolution of stellar systems from gas, we introduce various 
physical processes into the GRAPE-SPH code;
radiative cooling, star formation, feedback of SNe II, SNe Ia, and SWs, 
and chemical enrichment including the mass dependence of SNe II. 
Here we describe the formulation and the assumptions of each physical processes.

\subsubsection{Radiative Cooling}

Radiative cooling is modeled using an equilibrium cooling function.
If gas is primordial with no heavy elements ([Fe/H] $<-5$),
we compute the cooling rates using two-body processes of H and He, and free-free 
emission (Katz et al. 1996).
These processes are collisional excitation of neutral hydrogen (H$^0$) and singly 
ionized helium (He$^+$), collisional ionization of H$^0$, He$^0$, and He$^+$, standard 
recombination of H$^+$, He$^+$, and He$^{++}$, dielectric recombination of He$^+$, and 
free-free emission.
We compute a lookup table, which lists $({\cal{H}}-\Lambda)/n_{\rm H}^2$ as a function of 
temperature and density.
We then evaluate net cooling rates at intermediate values with cubic spline interpolation.

For metal-enriched gas ([Fe/H] $\ge -5$), we use a metallicity-dependent cooling function 
computed with the MAPPINGS III software by R.S. Sutherland (MAPPINGS III is the updates 
of MAPPINGS II that is described in \citealt{sut93}). 
The included processes are collisional line radiation, free-free and two-photon continuum, 
recombination, photoionization heating, collisional ionization, and Compton heating.
Heavy elements can significantly enhance the cooling rate.
At $T > 10^7$ K, Fe group line emission processes largely determine the cooling 
function, whereas at lower temperatures the lighter atoms such as C, O, and Ne dominate.
Although the metallicity effect is small at $T \sim 10^4$ K,
the cooling rate with [Fe/H] $=0$ is $\sim 100$ times larger than 
that for the primordial gas around $T \sim 10^5$ K.
The radiative cooling, and hence the star formation rate, strongly depends on the metallicity.
In the table of cooling function that we use,
cooling rates are given as functions of [Fe/H],
and the elemental abundance ratios are set to be constant for given [Fe/H] according to the relations in the solar neighborhood; Galactic halo stars (i.e., [O/Fe]$=0.5$) for [Fe/H] $\le -1$, and the solar values for [Fe/H] $\ge 0$, and interpolated the halo and solar values for $-1<$ [Fe/H] $<0$.
Since the elemental abundance ratios depend on the star formation history and the inhomogeneous mixing of SNe II and Ia, the cooling tables depending on [O/Fe] are required to increase the accuracy.

\subsubsection{Star Formation}

The treatment of star formation is similar to that in \citet{kat92},
which is widely used in GRAPE-SPH simulations.
If a gas particle satisfies the following star formation criteria,
a fractional part of the mass of the gas particle turns to a star particle.
Since an individual star particle has the mass of $10^{5-7} M_\odot$,
it dose not represent a single star, but an association of many stars.

Our star formation criteria are (1) converging flow, (2) rapid cooling, and (3) Jeans unstable on a particle scale;
\begin{eqnarray}
&\mbox{(1)}& (\nabla \cdot \mbox{\boldmath$v$})_i < 0  ,\\
&\mbox{(2)}& t_{\rm cool} < t_{\rm dyn}  ,\\
&\mbox{(3)}& t_{\rm dyn} < t_{\rm sound}  .
\end{eqnarray}
Here $t_{\rm cool}$, $t_{\rm dyn}$, and $t_{\rm sound}$ are the cooling time, the dynamical time of a particle, 
and the sound crossing time, respectively, and are expressed as
\begin{equation}
t_{\rm cool} = \frac{\rho u}{\Lambda}  ,
\end{equation}
\begin{equation}
t_{\rm dyn} = \frac{1}{\sqrt{4\pi G\rho}}  ,
\end{equation}
\begin{equation}
t_{\rm sound} = \frac{h_i}{c_{\rm s}}   ,
\end{equation}
where $\mu$ is the mean molecular weight, $u$ is the specific thermal energy, and $c_{\rm s}$ is the local sound speed.

We assume that the star formation timescale is proportional to the dynamical timescale;
\begin{equation}
\label{eq:c}
t_{\rm sf}=\frac{1}{c}t_{\rm dyn}  ,
\end{equation}
where the star formation parameter $c$ is set to be $1.0$.
With $c=0.1$, star formation takes place slowly, which results in too blue colors compared with the observation (see \S \ref{sec:discussion} for more discussion).
The star formation rate (SFR) is defined as
\begin{equation}
\frac{D\rho_*}{Dt} = -\frac{1}{t_{\rm sf}}\rho = -\frac{c}{t_{\rm dyn}}\rho = -c \sqrt{4\pi G}\rho^\frac{3}{2}  ,
\end{equation}
where $\rho_*$ is the stellar density.
In other words, we adopt the \citet{sch59} law, where
the SFR is the power of the gas fraction, and the power index is $n_{\rm s}=1.5$. 
This is consistent with the H$\alpha$ observation of disk galaxies that implies $n_{\rm s}=1.3\pm0.3$ (\citealt{ken89}).

The star formation criteria are estimated with the time interval of $\Delta t_{\rm sf}$, which is set to be $2$ Myr.
The star formation probability $P$ for a gas particle forming stars during $\Delta t$ is given by \citet{kat92} as
\begin{equation}
\label{eq:pro_cri}
P = 1 - \exp\left[-\frac{c}{t_{\rm dyn}}\Delta t_{\rm sf}\right]
=1 - \exp\left[-c\sqrt{4\pi G\rho}\Delta t_{\rm sf}\right]  .
\end{equation}
A random number between 0 and 1 is drawn to determine whether the gas particle forms stars 
during $\Delta t_{\rm sf}$: If $P$ is larger than the random number, star formation occurs.
$P$ is larger for higher density, and the typical density to form stars is higher with smaller $c$.
For $c=1.0$ and $c=0.1$, most stars form in the region with 
$\rho \gtsim 10^{-24}$ and $10^{-22}$ [g cm$^{-3}$], respectively.
Practically, the lower limit of the gas density is given by $c$.

When the above criteria (Eq.[6-7] and Eq.[14]) are satisfied, a part of material of the gas particle turns to a star particle, which newly forms near the gas particle.
We follow the scheme in \citet{nak03}, where
the initial mass of the star particle, $m_*^0$, is derived from the integration of 
the SFR over the time interval $\Delta t_{\rm sf}$;
\begin{equation}\label{eq:mstar}
m_*^0 = \rho\pi h_i^3 \left(1-\exp\left[-\frac{c}{t_{\rm dyn}}\Delta t_{\rm sf} \right] \right)  .
\end{equation}

\subsubsection{Feedback}

The evolved stars eject surrounding materials and heavy elements via stellar winds 
and supernova explosions. Those heat up, accelerate, and enrich the circumstellar and interstellar medium.
High energy explosions like supernovae produce high temperature and low density 
regions in the interstellar medium. In SPH methods, the numerical accuracy for high 
density regions is much better than in mesh based methods, but the accuracy for low density regions is poorer. 
In an SPH simulation such as those of \citet{nav93}, 
the numerical resolution (100-1000 pc) is larger than the typical size of supernova 
remnants (several tens of pc). 
Thus, because of the nature of the SPH method and the lack of resolution in current 
computing resources, it is necessary to simplify the release of the energy, momentum, and mass from stars.
We therefore assume that energy and heavy elements that are ejected from a star particle are 
equally distributed to the surrounding gas particles within a sphere of feedback radius 
$r_{\rm f}$, which is a parameter that controls the mixing of heavy elements. We set 
$r_{\rm f}=1$ kpc, which gives a good fit to the chemical evolution
in the Milky Way Galaxy (\citealt{kobD02}).

In this paper, we distribute the feedback energy in purely thermal form.
\citet{nav93} proposed that the energy produced by a supernova explosion is 
distributed to neighbor gas particles of the star particle mostly as a thermal energy 
and the rest is distributed as a velocity perturbation to the gas particles; the fraction 
of energy in kinetic form is given by a free parameter $f_{\rm kin}$.
With $f_{\rm kin}>0$, star formation efficiency is smaller, surface brightness decreases at the center, and metal-rich gas blow out. 
If we adopt $f_{\rm kin}=0.1$, the effective radius is too large and the metallicity gradient is too shallow (see Fig.\ref{fig:param} in \S \ref{sec:discussion}).

The energy ejection rate $E_e$ from a star particle as a function of age $t$ is
\begin{equation}
E_e(t) = m_*^0 \left( e_{e,{\rm SW}}{\cal{R}}_{\rm SW}(t)+e_{e,{\rm II}}{\cal{R}}_{\rm II}(t)+e_{e,{\rm Ia}}{\cal{R}}_{\rm Ia}(t) \right)  .
\end{equation}
${\cal{R}}_{\rm SW}$, ${\cal{R}}_{\rm II}$, and ${\cal{R}}_{\rm Ia}$ are the rates of SWs, 
SNe II, and SNe Ia, respectively, of which formulations are described in \S \ref{sec:chem}.

The energy of SWs from solar metallicity stars 
is estimated to be typically $0.2\times10^{51}$ erg 
from the observation of OB associations (\citealt{abb82}).
Although there should be a mass dependence where massive stars eject larger energies, we adopted the typical value because of lack of observation.
The chemical abundance of the star significantly affects SWs as $\dot{M} \propto Z^{0.8}$ (\citealt{lei92}), 
thus we include the metallicity effect for very massive stars;
\begin{eqnarray}
e_{e,{\rm SW}} = \!\left\{\!
\begin{array}{ll}
0.2 \times 10^{51} \left(\frac{Z}{Z_\odot}\right)^{0.8} \!\!\!\!\!\! & (m_{2,u} <m\le m_u)\\
0.2 \times 10^{51} & (m_{2,\ell} <m\le m_{2,u})
\end{array}
\right. \!\!\! \mbox{[erg]}  ,
\end{eqnarray}
where $Z$ is the metallicity of the star particle
(see \S \ref{sec:chem} for the upper and lower mass limits).

Since there are supernovae with 10 times larger energy than typical supernovae, hypernovae, 
the energy of all supernovae should not be the same. 
However, since the energy distribution function of supernovae has not been established, we adopt typical values of
\begin{equation}
e_{e,{\rm II}} = 1.4 \times 10^{51} ~~(m_{2,\ell} <m\le m_{2,u})~~~\mbox{[erg]}
\end{equation}
and
\begin{equation}
e_{e,{\rm Ia}} = 1.3 \times 10^{51} ~~(m_{{\rm 1d},\ell} <m\le m_{{\rm 1d},u})~~~\mbox{[erg]}
\end{equation}
for SNe II (\citealt{bli00}) and SNe Ia (\citealt{nom84}), respectively
(see \S \ref{sec:chem} for the upper and lower mass limits).

\subsubsection{Chemical Enrichment}
\label{sec:chem}

A star particle is not a single star but an association of many stars.
We assume that a star particle is in fact a simple stellar population, 
which is defined as a single generation of coeval and chemically 
homogeneous stars of various masses, i.e., 
it consists of a number of stars with various masses but the same age and metallicity.
The mass, mass of heavy elements, and the spectral energy distribution of the star particle evolve as massive stars die.
From dying stars, gas is ejected into interstellar medium
by SWs, SNe II, and SNe Ia.

The mass of the stars associated with each star particle
is distributed according to an initial mass function (IMF).
The IMF is assumed to be invariant to time and metallicity as
\begin{equation}
\label{eq:imf}
\phi(m) \propto m^{-x}  ,
\end{equation}
which is normalized to unity at $m_\ell \leq m \leq m_u$. 
Theoretical arguments
indicate that the IMF originates from fragmentation of a gas cloud
almost independently of local physics in the gas
(\citealt{low76}; \citealt{sil77}).
In the solar neighborhood, the Salpeter IMF with $x=1.35$ (\citealt{sal55})
is a good approximation to the star counts, and gives a good fit to many properties of disk galaxies (\citealt{ken83}).
For ellipticals, the flatter IMF is favored to explain the red colors of giant ellipticals (\citealt{kodD97}).
We then adopt $x=1.10$ 
with the upper and lower masses of
$m_\ell=0.05 M_\odot$ and $m_u=120 M_\odot$.

The ejection rates of the mass and heavy element $i$ ($E_m$ and $E_{z_i}$) from the star particle are expressed as
\begin{equation}
E_m(t) = m_*^0 \left( e_{m,{\rm SW}}(t)+e_{m,{\rm II}}(t)+e_{m,{\rm Ia}}(t) \right)  ,
\end{equation}
and
\begin{equation}
E_{z_i}(t) = m_*^0 \left( e_{z_i,{\rm SW}}(t)+e_{z_i,{\rm II}}(t)+e_{z_i,{\rm Ia}}(t) \right)  .
\end{equation}
The ejection rates per mass are given by the following equations;
\begin{equation}
e_{m,{\rm SW}}=\left(\frac{Z}{Z_\odot}\right)^{0.8} 
\int_{\max[m_{{\rm 2},u},\,m_t]}^{m_u}\!\!\!\!\!\!\!\!\!\!\!\!\!\!\!(1-w_m)\,\phi(m)~dm  ,
\end{equation}
\begin{equation}
e_{m,{\rm II}}=\int_{m_t}^{m_{2,u}}\,(1-w_m)\,\phi(m)~dm  ,
\end{equation}
and
\begin{equation}
e_{m,{\rm Ia}}=m_{\rm CO}\,{\cal R}_{\rm Ia}(t) .
\end{equation}
Time dependence is in the lower mass limit for integrals,
the turn-off mass $m_t$, which
is the mass of the star with the main-sequence lifetime $\tau_m=t$.
For simplicity, the lifetime is determined as (\citealt{dav90})
\begin{equation}
\log \tau_m = 10.0+(-3.42+0.88\log m) \log m  ,
\end{equation}
which gives a little ($1.5$ times) longer lifetime for $m \gtsim 20 M_\odot$ compared 
with the metallicity-dependent lifetime of Kodama \& Arimoto (1997).
In SWs and SNe II, stars eject the envelope materials outside the remnants.
$w_m$ is the remnant mass fraction, 
which is the mass fraction of a black hole, a neutron star, or a white dwarf,
depending on the initial mass $m$.
For SNe Ia, all of the evolved He core (i.e., C+O white dwarfs) is ejected, and the mass of the white dwarf 
at the SN Ia explosion is $m_{\rm CO}=1.38 M_\odot$.

Heavy elements are also ejected at the following rates;
\begin{equation}
e_{z_i,{\rm SW}}=\left(\frac{Z}{Z_\odot}\right)^{0.8} \!\!
\int_{\max[m_{{\rm 2},u},\,m_t]}^{m_u}\!\!\! \!\!\!\!\!\!\!\!\!\!\!\!\!\!\!(1-w_m-p_{z_im,{\rm II}})\,Z_i\,
\phi(m)~dm  ,
\end{equation}
\begin{eqnarray}
e_{z_i,{\rm II}}&=&\int_{\max[m_{{\rm 2},\ell},\,m_t]}^{m_{2,u}}\!\!\!\!\!\!\!\!\!\!\!\!\!\!\!p_{z_im,{\rm II}}\,\phi(m)~dm \\
&+& \int_{\max[m_{{\rm 2},\ell},\,m_t]}^{m_{2,u}}\!\!\!\!\!\!\!\!\!\!\!\!\!\!\!(1-w_m-p_{z_im,{\rm II}})\,Z_i\,
\phi(m)~dm  ,
\end{eqnarray}
and
\begin{equation}
e_{z_i,{\rm Ia}}=m_{\rm CO}\,p_{z_im,{\rm Ia}}\,{\cal R}_{\rm Ia}(t) .
\end{equation}
In SWs and SNe II, the heavy elements in the envelope are ejected.
In SNe II and SNe Ia, the explosive nucleosynthesis takes place,
and heavy elements are newly produced.
$p_{z_im,{\rm II}}$ and $p_{z_im,{\rm Ia}}$ are the stellar yields,
which are the mass fractions
of newly produced and ejected heavy elements $i$, and are given from 
the supernovae nucleosynthesis model (Nomoto et al. 1997ab)
with $p_{z_im,{\rm II}}=0$ for $m<10 M_\odot$.
The upper and lower limits of SNe II are $m_{{\rm 2},u}=50 M_\odot$ and $m_{{\rm 2},\ell}=8 M_\odot$, respectively.
For stars with $50-120 M_\odot$, all mass of He core turns to a black hole.
Although some metals such as carbon are produced in SWs,
we here neglect them
because their contribution is much smaller than that of supernovae.
The dependence of $w_m$,
$p_{z_im,{\rm II}}$ and $p_{z_im,{\rm Ia}}$ on the stellar metallicity
has not been included.

The rates of SWs and SNe II, ${\cal R_{\rm SW}}$ and ${\cal R_{\rm II}}$, are obtained as
\begin{equation}
{\cal R}_{\rm SW}=\int_{\max[m_{{\rm 2},u},\,m_t]}^{m_u}\,
\frac{1}{m}\,\phi(m)~dm  ,
\end{equation}
and
\begin{equation}
{\cal R}_{\rm II}=\int_{\max[m_{{\rm 2},\ell},\,m_t]}^{m_{2,u}}\,
\frac{1}{m}\,\phi(m)~dm  .
\end{equation}
For the SN Ia rate ${\cal R_{\rm Ia}}$, there are several alternative role debate. 
Here we adopt an SN Ia model based on the single degenerate scenario including 
metallicity effects (\citealt{kob98}; \citealt{kob00});
\begin{eqnarray}
{\cal R}_{\rm Ia}&=&\!\!b~
\int_{\max[m_{{\rm 1p},\ell},\,m_t]}^{m_{{\rm 1p},u}}\,
\frac{1}{m}\,\phi(m)~dm \\
&\times&
\int_{\max[m_{{\rm 1d},\ell},\,m_t]}^{m_{{\rm 1d},u}}\,
\frac{1}{m}\,\phi_{\rm d}(m)~dm .
\label{eq:snia}
\end{eqnarray}
The primary star is a C+O white dwarf (WD) formed from a stars with initial mass between 
$m_{{\rm 1p},\ell}=3 M_\odot$ and $m_{{\rm 1p},u}=8 M_\odot=m_{{\rm 2},\ell}$.
This SN Ia scenario has
two types of secondary stars, main-sequence (MS) and red-giant (RG) stars.
We calculate the SN Ia rate for each binary system
(i.e., the MS+WD and the RG+WD system)
with respective $b$, $m_{{\rm 1d},\ell}$, and $m_{{\rm 1d},u}$, 
and combine them.
The mass ranges of the companion stars are given by the simulation of binary evolution, 
and are $0.9-1.5 M_\odot$ for the RG+WD system and $1.8-2.6 M_\odot$ for the MS+WD system.
Thus the lifetime of SNe Ia is $\sim 2-20$ Gyr and $0.5-1.5$ Gyr, respectively.
$b$ is the binary parameters, which is the fraction of white dwarfs that eventually 
produce SNe Ia, and we adopt the same value which are determined from the chemical 
evolution in the solar neighborhood; $[b_{\rm RG}, b_{\rm MS}]=[0.02, 0.05]$ (\citealt{kob00}).

The photometric evolution of a star particle is identical to the evolution of the simple stellar population, 
of which spectra $f_\lambda$ are taken from Kodama \& Arimoto (1997) as a function of age $t$ and metallicity $Z$.
The absolute magnitude is calculated by using passbands to the the photometric 
system, which are the same as Kodama (1997).

\subsection{Initial Condition}
\label{sec:init}

A cosmological initial condition is generated with the following three steps.
Throughout the paper, we set the cosmological parameters of
$H_0=50$ km s$^{-1}$ Mpc$^{-1}$, $\Omega_m=1.0$, $\Omega_\Lambda=0$, and $\sigma_8=1.0$. (In the following, lengths and masses are for $H_0=50$.)
First, we generate a periodic boundary condition with the lattice size $5$ Mpc having a top-hat perturbation of amplitude $1\sigma$ or $3\sigma$ in radius $1.25$ Mpc by using the COSMICS package (\citealt{ber95}). 
This package uses the standard Zel'dovich approximation (\citealt{zel70}; \citealt{efs85}) 
to compute the displacements and velocities of the dark matter particles from a Gaussian 
random density field. The power spectrum of this density field represents the CDM spectrum for the above cosmological parameters.
To generate constrained density field, a path integral method is used (\citealt{ber87}).
We set the parameters in COSMICS to get the density field realization with the desired number, 
mass, and the starting redshift of $z \sim 25$.
By changing the seed for the random number generator, we obtain a different density field.
Giving the seed at random, we obtain a random sample of initial conditions.

Second, we pick up particles in a spherical region with the comoving radius of $\sim 1.5$ Mpc, the mass of $\sim 10^{12} M_\odot$ (baryon fraction of $0.1$), 
and $N_{\rm tot}$ particles (the half for gas and the rest for dark matter).
In this paper, we set two different resolutions; $N_{\rm tot} \sim 10000$ and $60000$.
The mass of a dark matter particle is $\sim 1.8 \times 10^8 M_\odot$ and $\sim 3.0 \times 10^7 M_\odot$, and the mass of 
a gas particle is $\sim 2.0 \times 10^7 M_\odot$ and $\sim 3.3 \times 10^6 M_\odot$, respectively.

Third, we give the initial angular momentum to the system in rigid rotation
because the simulated field is not enough large to generate tidal torque.
The typical spin parameter of a virialized halo in CDM cosmology ranges from $0.01$ to $0.1$ and the median is $0.05$ according to the numerical simulations 
(e.g., \citealt{war92}).
We adopt the constant spin parameter $\lambda$ as small as $\sim 0.02$.
With larger $\lambda$, a spiral galaxy form if the galaxy does not undergo a major merger.
If a major merger occurs, an elliptical galaxy form, even if the spin parameter is initially set to be as large as $\lambda \sim 0.1$,
Thus, we should note that such ellipticals are not included in our sample (see \S \ref{sec:discussion} for more discussion).
We also add the corresponding Hubble velocity to the velocity field of the sphere since 
the equation of motion is integrated not in comoving units but in physical units.

This scheme is similar to that in \citet{kat92}, and has been used in many works (e.g., \citealt{ste94}; \citealt{nak03}; \citealt{kaw03}).
Simulations from true CDM initial conditions excluding the artificial boundary effect were firstly carried out by \citet{nav94}.
We should keep in mind the boundary effect of our initial condition;
materials in the outside of the simulated sphere is neglected, which makes the artificial cut off of the mass accretion
(see \S \ref{sec:discussion} for more discussion).

\section{RESULTS}

\subsection{EVOLUTION HISTORIES}
\label{sec:class}

\begin{figure*}
\begin{center}
\includegraphics[height=17cm,angle=-90]{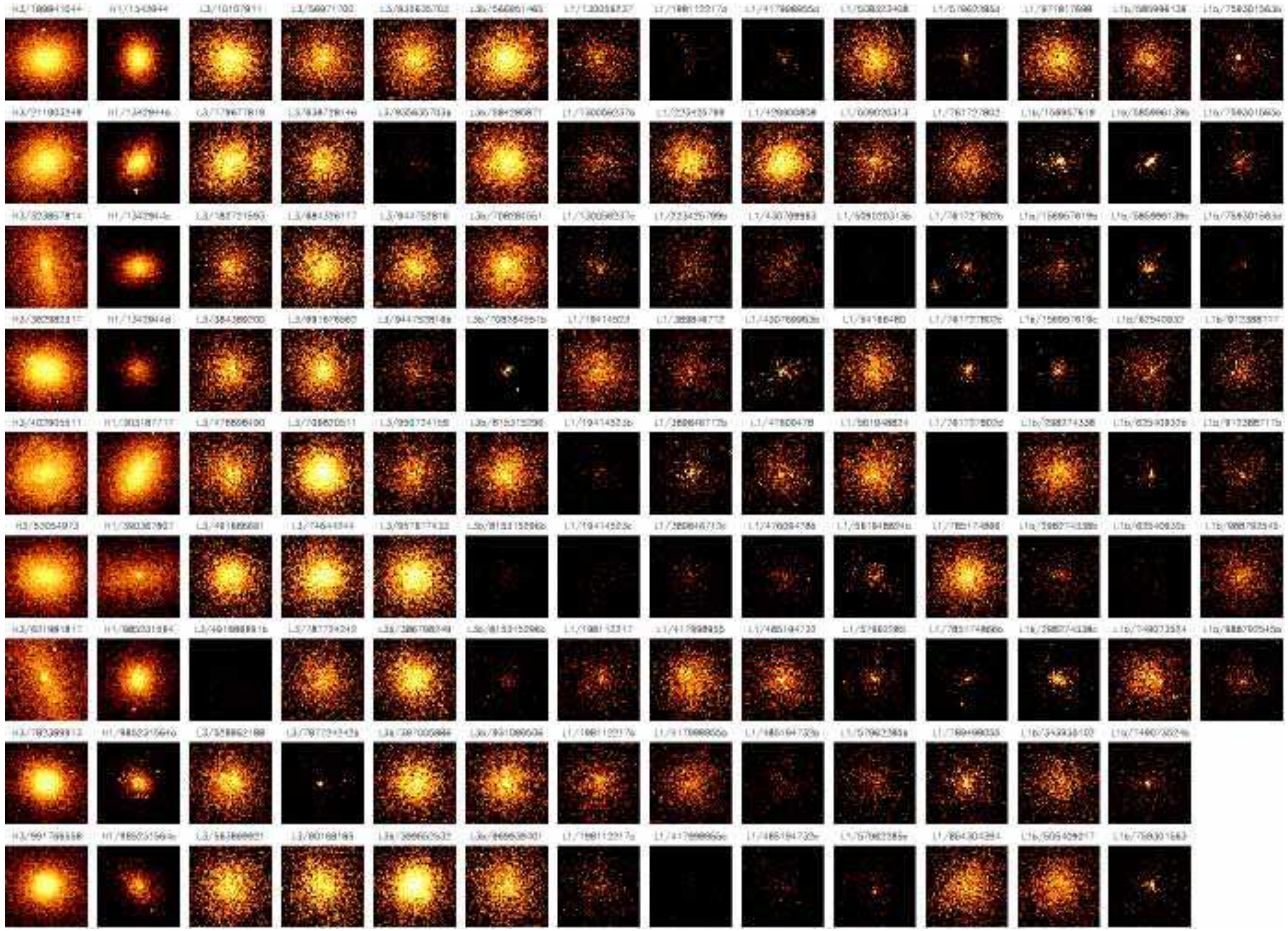}
\caption{\label{fig:map}
V-luminosity distributions in $\pm 10$ kpc of 124 simulated galaxies with the stellar masses of $\sim 10^{8-11} M_\odot$.
The number on the top of each image is ID number in the simulation.
}
\end{center}
\end{figure*}

We simulate the chemodynamical evolution of 72 fields with different cosmological initial 
conditions (see \S \ref{sec:init}). 
It takes $\sim10$ days to simulate one field  with high-resolution (the number of particles 
$N\sim60000$), and $1$ day with low-resolution ($N\sim10000$).
By the present time (i.e., $t=13.2$ Gyr, $z=0$), 
in 42 cases one galaxy forms in the center of the field, 
in 16 cases one galaxy and several subgalaxies, and 
in 14 cases a few galaxies with comparable masses.
We select galaxies having stellar masses in a 20 kpc sphere larger than $4.5 \times 10^7 M_\odot$.
Although many less-massive subgalaxies form, we discard them because our resolution 
is not enough to study them in detail.
We summarize the number of runs and the resulting galaxies in Table 1. 
\begin{table}
\caption{Number of simulated galaxies}
\label{tab:number}
\begin{tabular}{lccc}
\hline
\footnotesize
 & run & ellipticals & dwarfs \\ 
\hline
high resolution ($N\sim60000$) & 13 & 18 & 0 \\
low resolution ($N\sim10000$)  & 59 & 60 & 46 \\
total                          & 72 & 78 & 46 \\
\hline
\end{tabular}
\end{table}

Figure \ref{fig:map} shows the V-luminosity distributions in $\pm 10$ kpc of 124 simulated 
galaxies.
Luminosities are projected on the $X$-$Y$ plane (the rotational axis is $Z$), and smoothed over $0.3$ kpc.
The number on the top of each image is the ID number in our simulation.
The galaxies with the same number are in the same simulated field.
The character ``H'' and ``L'' respectively denotes the high- and low- resolution, and the 
following ``1'' and ``3'' respectively denotes the $1\sigma$ and $3\sigma$ over-dense regions.
The 18 galaxies in the left side are obtained with high-resolution simulations, which clearly show triaxial distributions.
The stellar masses span in the range $\sim 10^{8-11} M_\odot$, and cD galaxies are not included in our sample.
Among the 124 galaxies, 78 bright galaxies are elliptical galaxies, which have de Vaucouleurs' surface brightness profiles.
Since the initial angular momentum is set to be small (spin parameter $\lambda \sim 0.02$), no spirals form.
The remaining 46 faint galaxies are either dwarf ellipticals or dwarf irregular galaxies, all of 
which have small stellar masses as $M\ltsim10^9 M_\odot$.
Some dwarf galaxies show bright central cores, while others are diffusely distributing stellar systems.

\begin{figure*}
\begin{center}
\includegraphics[height=15cm,angle=-90]{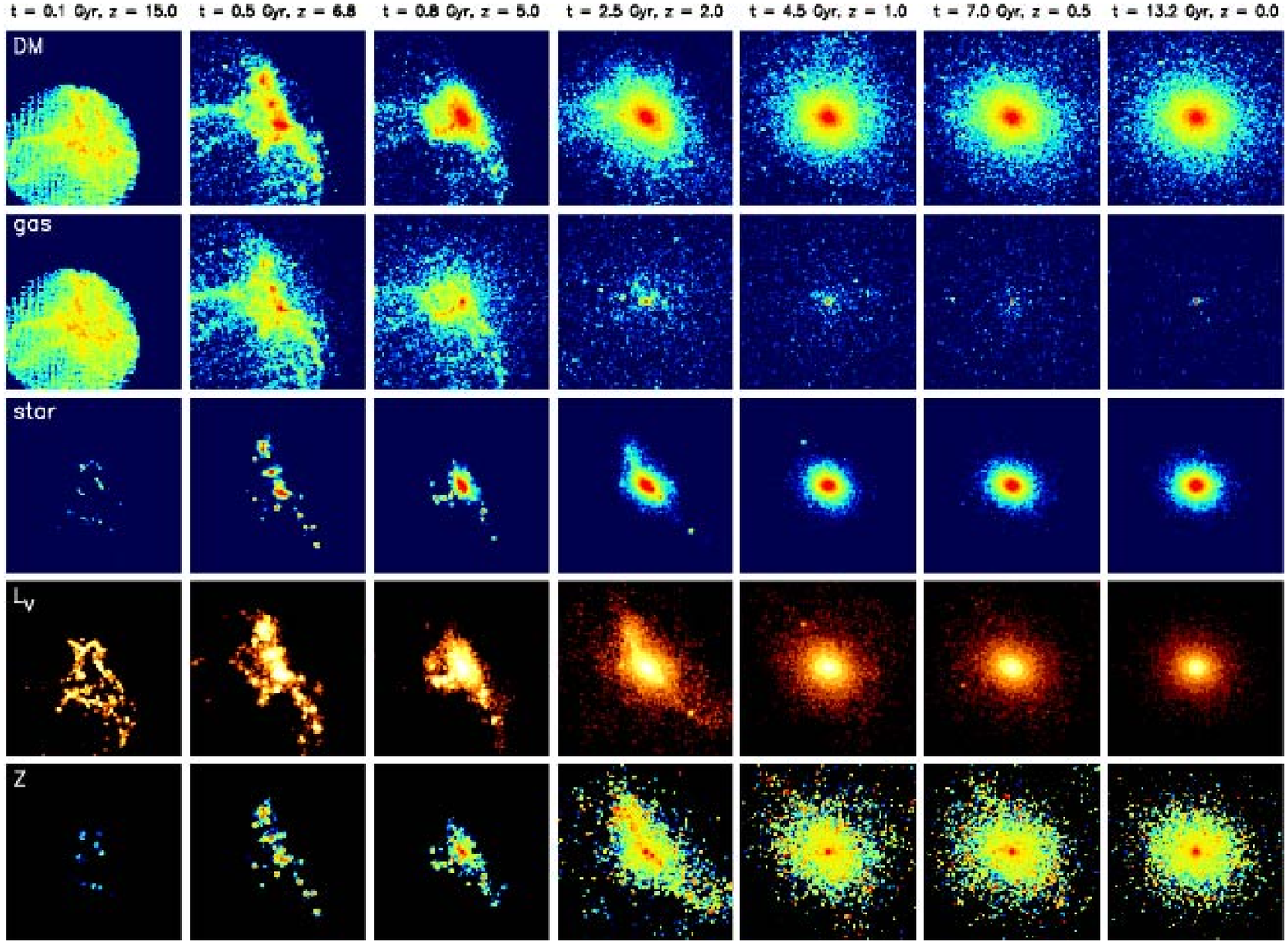}
\caption{\label{fig:evomapD}
The evolutions in $\pm 100$ kpc of dark matter (first lines), gas (second lines), 
stars (third lines), V-luminosity (forth lines), and stellar metallicity (fifth lines) of the galaxy that forms monolithically.
The metallicity range is $\log Z/Z_\odot=-1$ to $0.4$.
}
\end{center}
\end{figure*}

\begin{figure*}
\begin{center}
\includegraphics[height=15cm,angle=-90]{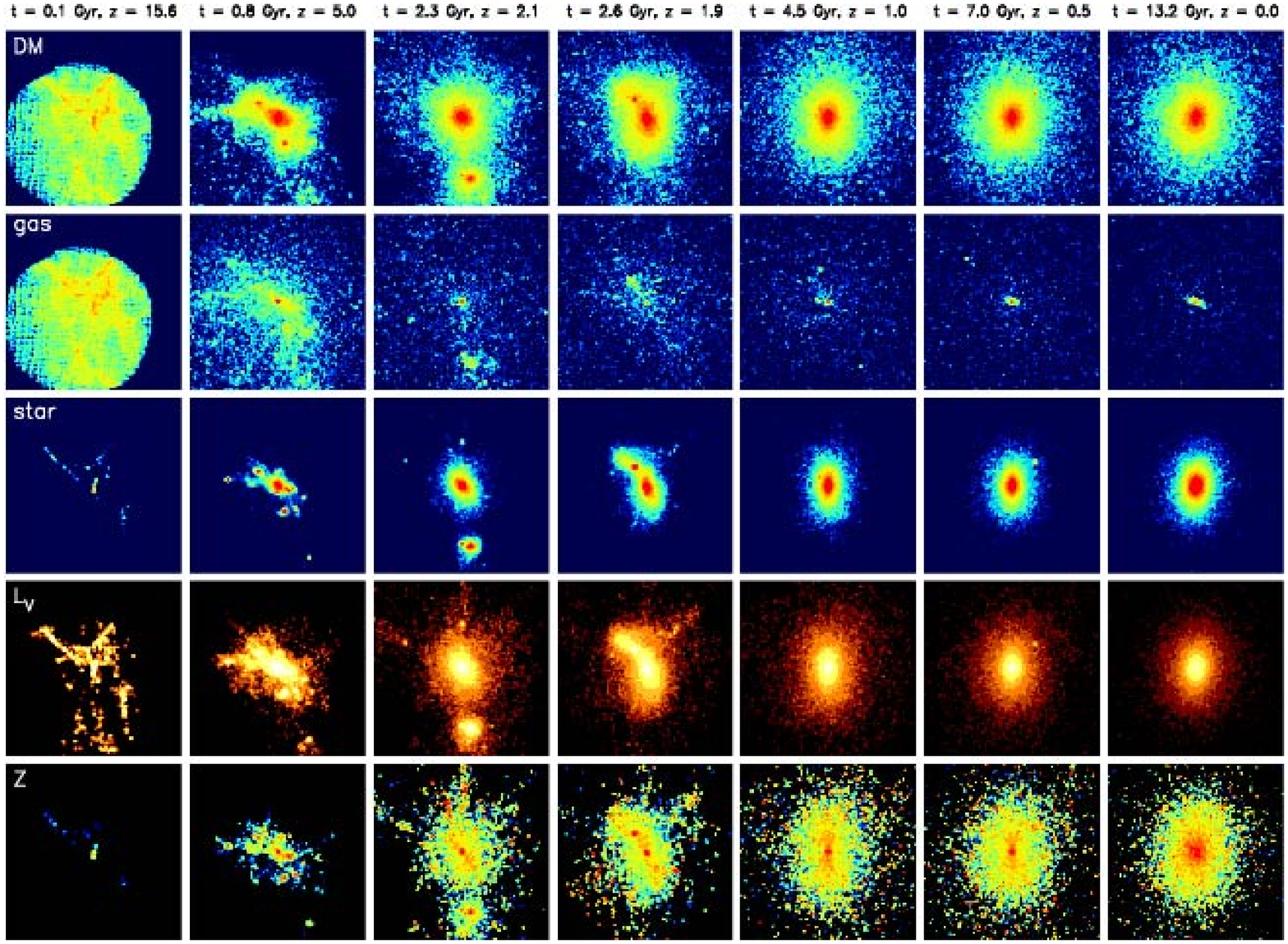}
\caption{\label{fig:evomapM}
The same as Figure \ref{fig:evomapD}, but for the galaxy that undergoes a major merger at $z \simeq 2.0$. 
}
\vspace*{-1cm}
\end{center}
\end{figure*}

Different galaxies undergo different evolution histories.
The difference is seeded in the initial condition.
Galaxies form through the successive merging of subgalaxies with various masses, which varies between
a major merger at one extreme and a monolithic collapse of slowly-rotating gas cloud at the other.
We show examples of the two cases from the high-resolution simulations in Figures \ref{fig:evomapD} and \ref{fig:evomapM}, which show the 
evolution in $\pm 100$ kpc on the $X$-$Z$ plane of 
dark matter (first lines), gas (second lines), 
stars (third lines), V-luminosity (forth lines), luminosity weighted stellar metallicity (fifth lines) of galaxies that form through monolithic collapse and through 
major merger, respectively.

Figure \ref{fig:evomapD} shows the evolution of a galaxy (ID H3/782389913) that forms monolithically.
At the beginning, the system expands according to the Hubble flow.
The CDM initial fluctuation produces the structures of nodes and filaments.
Gas cores form in the nodes, and stars form in the gas cores.
The surface brightness is as high as $14$ mag arcsec$^{-2}$ in rest-frame V-band.
Gas rich subgalaxies merge with one another, and the protogalaxy coalesces at $z\gtsim 3$. The accretion of 
small subgalaxies continues till $z\sim 2$, and after this no significant event happens.
There is only small amount of star formation at $z\ltsim0.5$, and 
the luminosity of the galaxy decreases gradually toward $z=0$.

Figure \ref{fig:evomapM} shows the evolution of a galaxy (ID H3/402905511) that undergoes a major merger at $z \simeq 2.0$. 
The primary galaxy forms in a large gas core at $z\gtsim 3$.
The secondary galaxy forms at the same time about $200$ kpc away, which comes to the center 
because of gravity, and merges with the primary galaxy. 
The secondary initially passes through the primary ($z=1.9$), and then oscillates decreasing in mass until the system is relaxed at $z=0.9$.
We should note that the edge of the dark matter distribution can be seen to fall onto the central object at $z \sim 6.8$ (Fig.\ref{fig:evomapD}) and $5.0$ (Fig.\ref{fig:evomapM}). This means that the mass accretion after then may be underestimated.

\begin{figure*}
\begin{center}
\includegraphics[height=17cm,angle=-90]{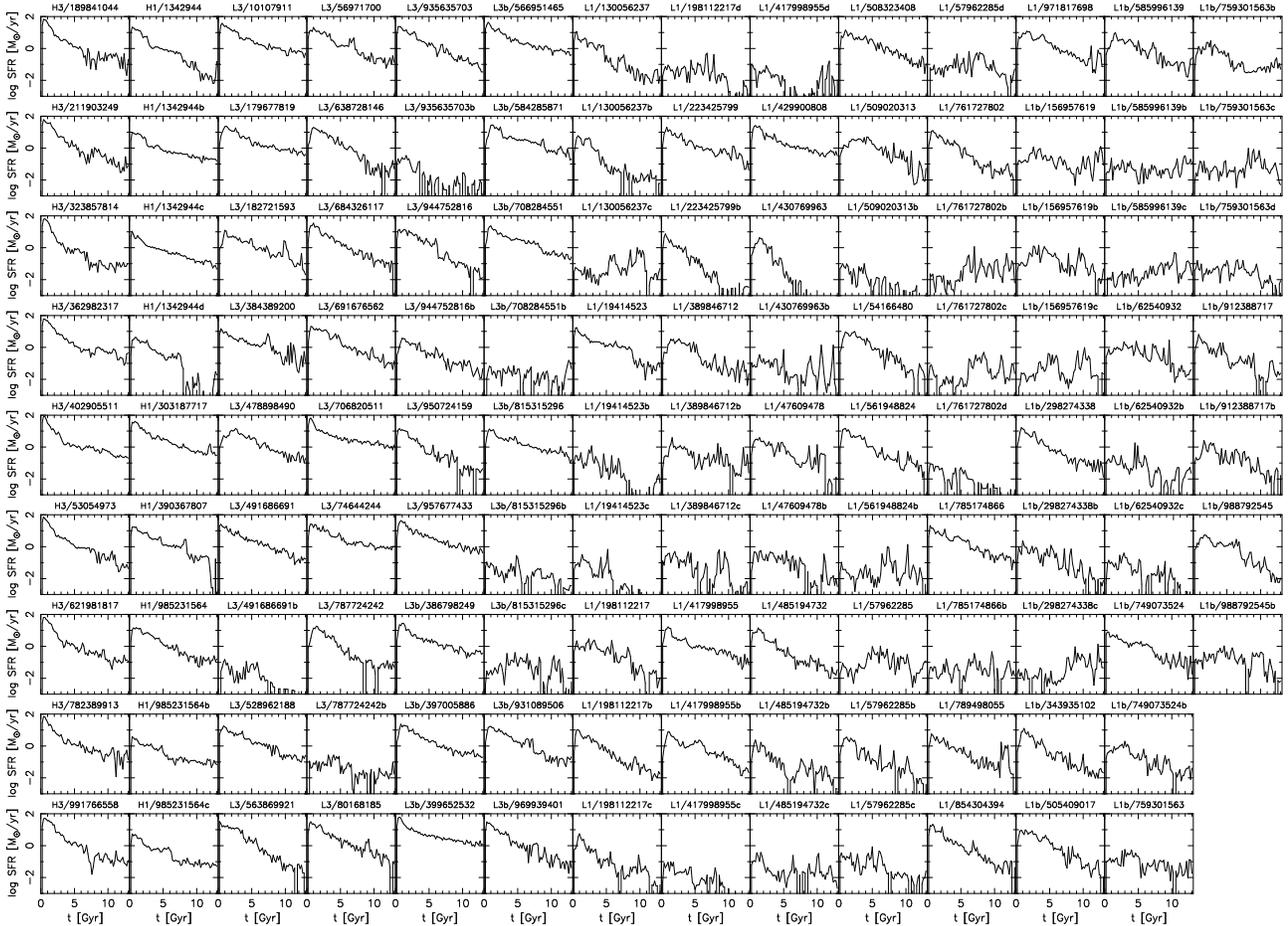}
\caption{\label{fig:sfr}
The star formation rates ($\log$ SFR [$M_\odot$ yr$^{-1}$]) as functions of time $t$ [Gyr] in a present-day galaxies ($r \le 20$ kpc and $|Z| \le 100$ kpc).
}
\end{center}
\end{figure*}

\begin{figure*}
\begin{center}
\includegraphics[height=17cm,angle=-90]{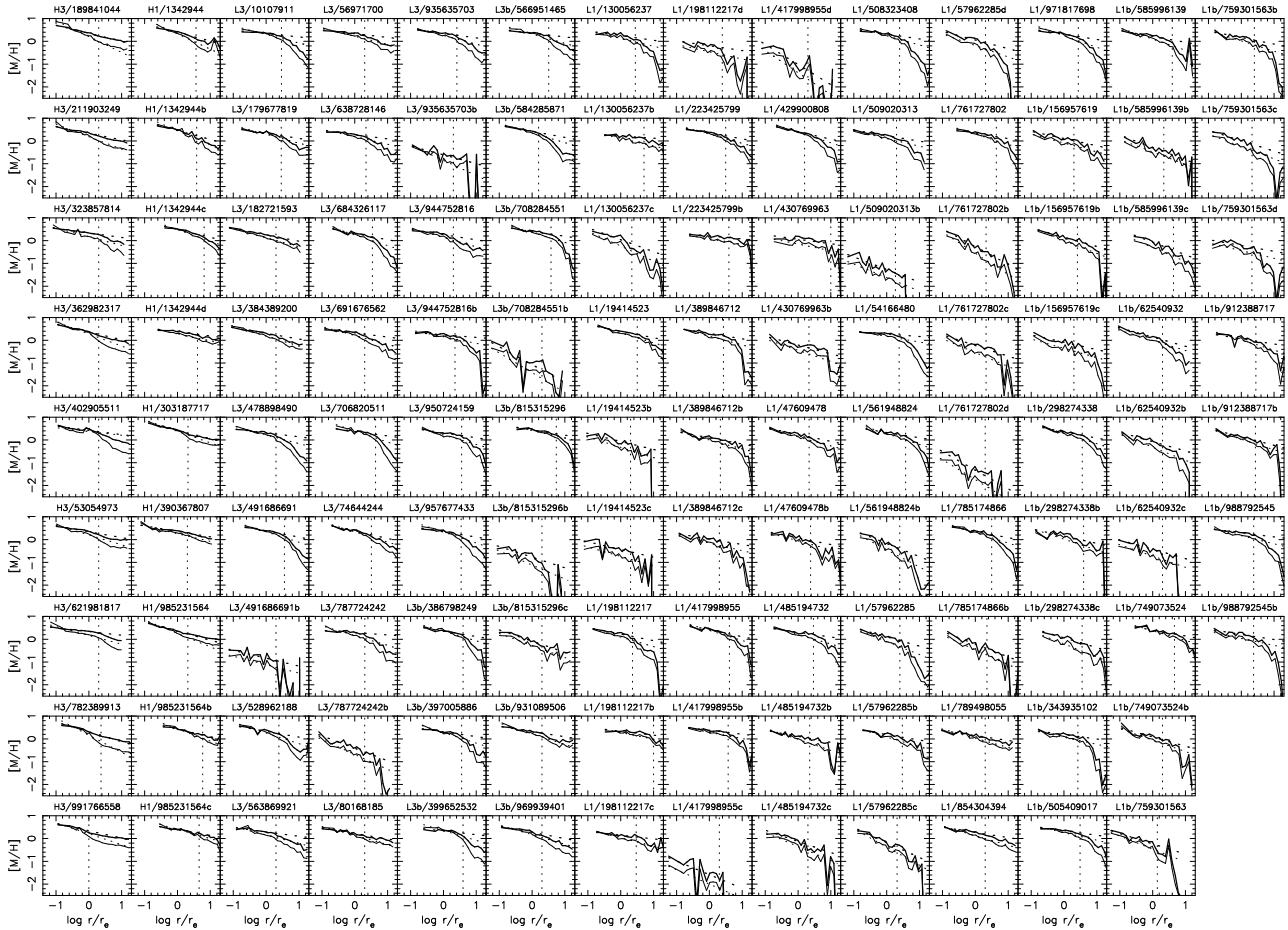}
\caption{\label{fig:zg}
The metallicity gradients, where [M/H] is plotted against $\log r/r_{\rm e}$.
The thick and thin lines show the oxygen and iron abundance gradient, respectively.
The fitting lines are shown with the dotted lines. 
The dotted vertical lines show the maximum radii used in the fitting.
}
\end{center}
\end{figure*}

We classify elliptical galaxies into the following 5 classes according to their merging histories.
\begin{enumerate}
\renewcommand{\labelenumi}{[\arabic{enumi}]}
\item {\bf Monolithic}--- Galaxies form through the assembly of many ($\gtsim 10$) gas-rich subgalaxies 
with the stellar masses of $M \sim 10^9 M_\odot$.
Such assembly generally has finished by $z \sim 3$, at least by $z \sim 2$. 
The gas fractions (i.e., the mass ratio of gas to baryon $f_{\rm g}\equiv M_{\rm g}/(M+M_{\rm g})$) of 
merging galaxies are larger than $0.5$.
The material of subgalaxies quietly accretes on the central galaxy.
It is difficult to discriminate these subgalaxies, and this assembly looks like a monolithic collapse.
\item {\bf Assembly}--- Galaxies form through the assembly of subgalaxies with $M\sim 10^{10} M_\odot$. 
The gas fractions $f_{\rm g}$ of subgalaxies are as large as $0.4$. 
Each subgalaxy has an evolved core, which violently merges with the others. 
While the subgalaxy passes through the central galaxy many times, many stars of the galaxy are 
stripped and some of them accrete on again.
\item {\bf Minor merger}--- The formation of the main component of the present-day galaxy is the 
same as above two classes, but these galaxies undergo minor merger events at $z \ltsim 3$.
We define the minor merger when the stellar mass ratio of the merging galaxies 
$f \equiv M_2/M_1$ ($M_1 \ge M_2$) ranges from $\sim 0.01$ to $\sim 0.2$.
With such minor merger events, the surface brightness profile and the metallicity gradients are not affected so much.
\item {\bf Major merger}--- Galaxies undergo the major merger with $f \gtsim 0.2$ at $z \ltsim 3$.
The mass of the primary galaxy is no more than five times larger than the mass of the secondary galaxy.
The redshift $z \sim 3$ generally corresponds to the galaxy formation epoch, and the 
main component of the present-day galaxy forms at $z \gtsim 3$.
Thus, the major merger occurs after the most stars in the present-day galaxies formed.
The merger event destroys the metallicity gradient that has existed in pre-merger galaxy in a way 
depending on the mass ratio $f$ and the gas mass of the secondary galaxy (see \S \ref{sec:grad}).
\item {\bf Multiple major mergers}--- Galaxies undergo a major merger ($f \gtsim 0.2$) 
and one or two other mergers with $f \gtsim 0.1$. 
\end{enumerate}
Dwarf galaxies are classified into the following 4 classes according to their star formation histories. 
Observationally, the first class of galaxies are dwarf ellipticals, the others are dwarf irregulars. 
\begin{enumerate}
\renewcommand{\labelenumi}{[D\arabic{enumi}]}
\item {\bf Initial star burst}--- Galaxies form with the initial star burst at $z \gtsim 1$. 
In some galaxies, many supernova explosions occur and cause the galactic winds. 
In other galaxies, the gas is not ejected completely, but the gas density is so 
small that only few stars form at lower redshifts. Thus the colors are red, and these 
dwarfs follow the same color-magnitude relation as giant ellipticals.
\item {\bf Continuous star formation}--- Galaxies grow though continuous star formation. 
After the initial star burst, there are the accretion of gas clumps and/or the interaction 
with other galaxies, which make the star formation continue to lower redshifts.
\item {\bf Continuous star formation with recent star burst}--- The same as [D2], 
but a star burst occurs at the recent $2-3$ Gyr. Thus the galaxy colors are blue.
\item {\bf Recent star burst}--- Galaxies form through recent star bursts at $z\sim0.7$. 
Such star bursts are induced by gas accretion and/or galaxy interactions. 
The colors are blue, and it is impossible to distinguish [D3] and [D4] with colors alone.
\end{enumerate}
The number of galaxies in each class is [1] 5 (4.0\%), [2] 18 (15\%), [3] 19 (15\%), 
[4] 25 (20\%), [5] 11 (8.9\%), [D1] 20 (16\%), [D2] 13 (10\%), [D3] 9 (7.3\%), and [D4] 4 (3.2\%).
The percentages of non-major merger ([1]-[3]) and 
major merger galaxies ([4]-[5]) are 34\% and 29\%, respectively.
We should note that the number of merger galaxies tends to be underestimated because of the boundary effect of our initial condition.

Figure \ref{fig:sfr} shows the star formation rates ($\log$ SFR [$M_\odot$ yr$^{-1}$]) 
as a function of time $t$ [Gyr], which
are derived from the ages and masses of stars that belong to the galaxy at present. 
Thus, these rates are not for stars formed in the region that is defined as the galaxy at each redshift, 
but for stars that formed anywhere and are today part of the galaxy.
This definition of the SFR will be the same as in any observation that estimates the star formation 
history of a galaxy from its stellar populations.
As clearly shown,
all elliptical galaxies form with an initial star burst at $z \gtsim 2$, whereas dwarf galaxies 
undergo relatively continuous star formation.
The SFR decreases because the gas is exhausted in the galaxy.
The secondary star burst is induced by the accretion of gas clumps and/or the merging of gas-rich galaxies.
Not all merging events induce a secondary star burst; the fraction of merging events that 
induce such a star burst is about $10\%$.
For all ellipticals, the initial star burst is always larger than the secondary one.
The typical timescale of initial star burst is found to be $1-2$ Gyr, which is much longer than 
the $0.1$ Gyr that is commonly adopted in the one-zone model (e.g., \citealt{kod97}).
Such SFR is due to the artificial cut-off of mass accretion caused by our initial condition, but is required from the observation of ellipticals as summarized in \S \ref{sec:intro}.

We should note that star formation has not completely stopped at present in most elliptical 
galaxies. Galactic winds are hard to generate in our simulated giant ellipticals 
(see \S \ref{sec:discussion} for more discussion). 
At the center of the present-day galaxy, the dynamical potential is so deep that the gas 
density is high. In such regions, super metal-rich stars ($Z\sim10Z_\odot$) form in the simulation.
In several dwarf galaxies with $M\ltsim$$10^9 M_\odot$ (e.g., ID L1/223425799b and L1/430769963), 
weak galactic winds can be seen even if $f_{\rm kin}=0$.
However, enough gas is heated up and blows away gradually
by the input of thermal energy of supernovae.
The global gas fraction of the simulated fields spans over $50-90\%$, and
the gas fraction in a galaxy (a sphere of $2 r_{\rm e}$) is $1-10\%$ and $30-80\%$ 
for giants and dwarfs, respectively.
The fraction of heavy elements locked into stars in a galaxy spans $20-40\%$
which is consistent with the observation (e.g., \citealt{ren02}).

\begin{figure*}
\begin{center}
\includegraphics[width=16cm]{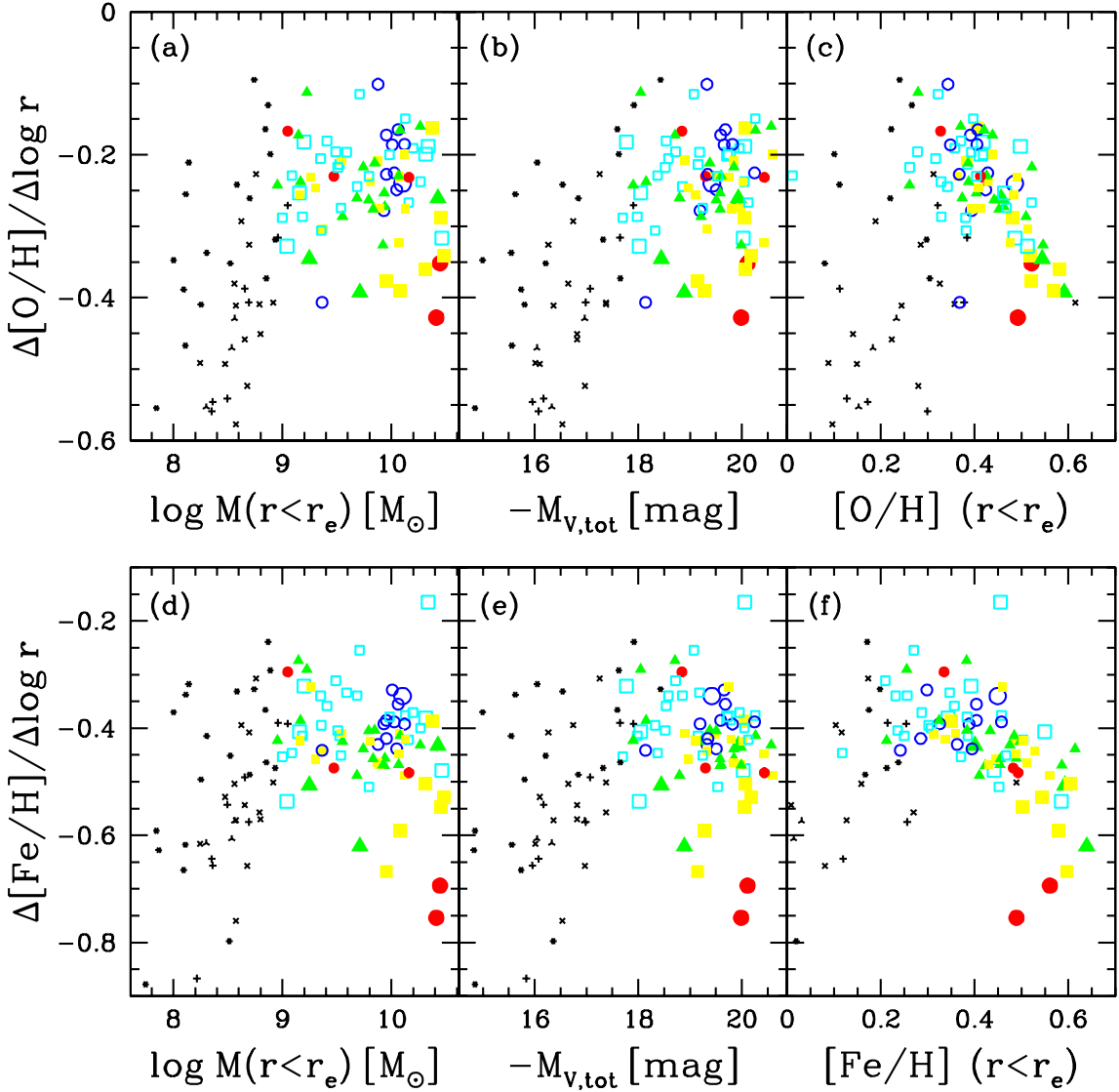}
\caption{\label{fig:grad}
Metallicity gradients versus (a,d) stellar masses in $r_{\rm e}$, 
(b,e) total luminosities, and (c,f) 
mean stellar metallicities in $r_{\rm e}$. The upper (a-c) and lower panels (d-f) 
show the gradients of oxygen and iron abundances, respectively.
All points are for the simulated galaxies, and the large points denote galaxies simulated with high-resolution.
The shape of symbols show the merging histories for elliptical galaxies and
star formation histories for dwarf galaxies; [1] monolithic 
(red filled circles), [2] assembly (yellow filled squares), [3] minor merger (green filled triangles), 
[4] major merger (cyan open squares), and [5] multiple major merger (blue open circles);
[D1] initial star burst (asterisks), [D2] 
continuous star formation (crosses), [D3] continuous star formation with recent star 
burst (plus), and [D4] recent star burst (three-pointed stars).
}
\end{center}
\end{figure*}

\subsection{METALLICITY GRADIENTS}
\label{sec:fit}

Galaxies are observed in projection on the sky. In order to compare the simulated results with 
observation, foreground and background particles should be excluded properly.
We define a galaxy as the projection of $|Z|\le100$ kpc on the $X$-$Y$ plane. Below radius $r$ means 
a projected radius.
We then derive the effective radius $r_{\rm e}$ by fitting a de Vaucouleurs' law to the surface brightness profile.
For the simulated galaxies, we exclude the central region with $r = 0-1$ kpc and $0-2$ kpc 
when fitting simulations with high- and low- resolution, respectively, because the 
surface brightness is smeared out due to the softening of the gravity.
In the outer region, to derive proper effective radii in the simulation, we use large enough 
regions such as $40-80$ kpc.
In some case that galaxies show a local excess in the surface brightness because of the existence of the 
satellite subgalaxies, we only use the inner part.

Observationally, metallicity gradients are derived from the line-strength gradients, and 
line-strength is converted to metallicity using an index-metallicity relation 
derived from spectral synthesis models (e.g., \citealt{kod97}).
Thus the observed metallicity is the luminosity weighted metallicity.
In the observational data,
the gradients are smeared out due to poor seeing conditions
at galaxy centers with ${\log r/r_{\rm e}} \ltsim -1.5$, 
and in the outer regions with $\gtsim 2 r_{\rm e}$ errors arising from the sky subtraction give poor fits. 
Kobayashi \& Arimoto (1999) excluded these regions from the fitting.
In the simulation, we provide the metallicity weighted by V-luminosity, 
because the absorption indices usually observed such as Mg$_2$ and Fe$_1$ are in the V-band.
Since we find no significant difference in the metallicity gradients at $r=1-2$ kpc for the high- and low-resolutions, 
the innermost boundary of the fitting region is set at $1$ kpc for both resolutions. 
This value is larger than the radius of ${\log r/r_{\rm e}} \sim -1.5$.
The outer boundary is set at $2 r_{\rm e}$ for most galaxies,
and at $10$ kpc for galaxies in the case with small $r_{\rm e}$.
The contribution of satellite galaxies is also excluded.

Figure \ref{fig:zg} shows the metallicity gradients of 124 galaxies, 
where [M/H] is plotted against $\log r/r_{\rm e}$.
The thick and thin lines show the oxygen and iron abundance gradient, respectively. 
The fitting lines are shown with the dotted lines.
The dotted vertical lines show the maximum radii used in the fitting. 
As is clearly seen, the metallicity gradients are various, some are steep and others flat. 

\subsection{RELATION AGAINST MASS?}
\label{sec:grad}

Figure \ref{fig:grad} shows the V-luminosity weighted metallicity gradients versus stellar 
mass within $r_{\rm e}$ (a and d), total luminosities derived from the de Vaucouleurs' law (b and e), 
and luminosity weighted mean stellar metallicities in $r_{\rm e}$ (c and f). The upper (a-c) and lower 
panels (d-f) show the gradients of oxygen and iron abundances, respectively.
All points are for the simulated galaxies, and the large points denote galaxies simulated with high-resolution.
The symbols show the merging histories for elliptical galaxies and star formation histories for 
dwarf galaxies; [1] monolithic (filled circles), 
[2] assembly (filled squares), [3] minor merger (filled triangles), [4] major merger (open squares), 
and [5] multiple major merger (open circles); [D1] initial star burst (asterisks), [D2] continuous 
star formation (crosses), [D3] continuous star formation with recent star burst (plus), and [D4] 
recent star burst (three-pointed stars).

The remarkable result is that there is no correlation between the gradients and masses or luminosities in Figure \ref{fig:grad}.
The lack of these relations has already been noticed in the observational data, 
and the origin of the scatter has been argued (\citealt{kob99}).
It seems that more metal-rich simulated ellipticals tend to have steeper 
gradients, 
but it is hard to find any relation if simulated dwarfs are included. 
The origin of scatter is clearly shown with the symbols;
the galaxies that form monolithically (filled circles and squares) 
have steeper gradients, and the galaxies that undergo major mergers (open circles and squares) 
have shallower gradients.
Therefore, we conclude that the metallicity gradients do not depend on the galaxy mass, and the 
variety of the gradients stems from the difference in the merging history.

\begin{figure}
\begin{center}
\includegraphics[width=8cm]{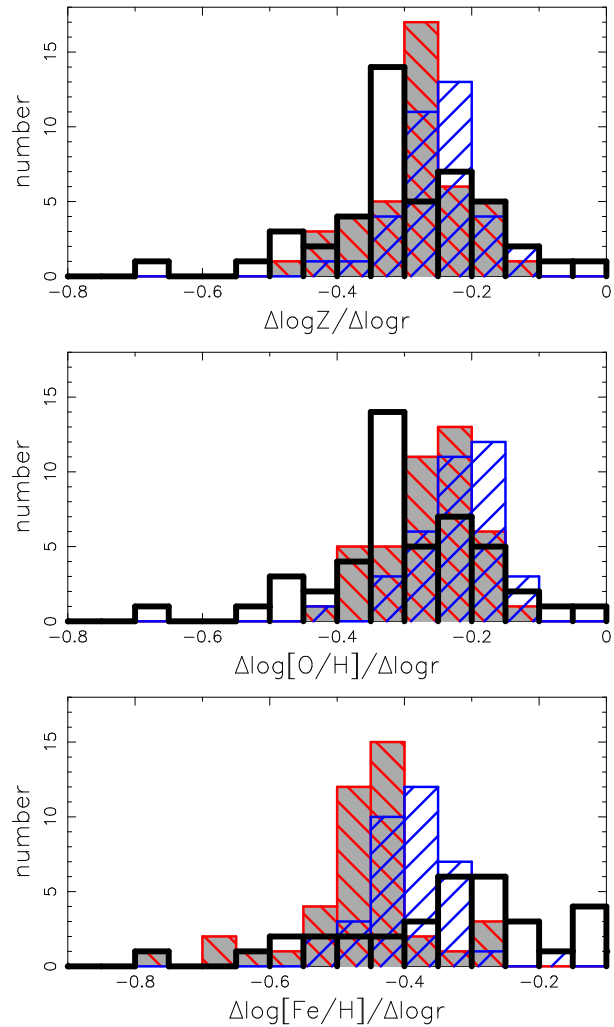}
\caption{\label{fig:gh}
The histograms of the gradients for metallicity $Z$ (upper panel), oxygen (middle panel), and iron abundance (lower panel).
The gray and hatched area are for non-major merger (red) and major merger (blue) galaxies, respectively.
The thick lines show the observation with Mg$_2$ (upper and middle panels) or  Fe$_1$ (lower panel) 
index from Kobayashi \& Arimoto (1999).
}
\end{center}
\end{figure}

The distributions of metallicity gradients for [1] monolithic collapse, [2] assembly, and [3] minor 
merger are very similar. 
The minor mergers do not affect so much the metallicity gradients.
The distributions for [4] major merger and [5] multiple major mergers are also similar. 
Then we joint the 5 classes into 2 large classes: [A] non-major merger galaxy including [1]-[3], 
and [B] major merger galaxy including [4] and [5].
Figure \ref{fig:gh} show the histograms of the metallicity gradients for the 2 classes using 
metallicity $Z$ (top panel), oxygen (middle panel), and iron abundance (bottom panel).
The distributions for [A] non-major merger (gray area) and [B] major merger (hatched area) 
galaxies are different.
The typical gradients for non-major merger and major merger galaxies 
are $\Delta\log Z/\Delta\log r \simeq -0.30$ and $-0.22$, $\Delta{\rm [O/H]}/\Delta\log r \simeq -0.30$ and $-0.18$, 
and $\Delta{\rm [Fe/H]}/\Delta\log r \simeq -0.45$ and $-0.38$, respectively.
The galaxies with gradients as steep as $\Delta\log Z/\Delta\log r \le -0.35$ are all non-major merger galaxies.

The thick lines show the observed metallicity gradient distributions for Mg$_2$ (top and middle panels) and Fe$_1$ (bottom panel) using data from Kobayashi \& Arimoto (1999).
(The reason to compare oxygen gradients with the Mg$_2$ observation is that the yields relative to solar value are almost the same between O and Mg.)
The simulated $Z$ gradients are consistent with the Mg$_2$ observation both with the mean value of $\Delta\log Z/\Delta\log r \simeq -0.3$ and the dispersion of $\pm 0.2$.
The oxygen gradients are shallower by $0.05$ dex than the Mg$_2$ observation.
This is because the Mg$_2$ index is not a good indicator for magnesium abundance, but rather for metallicity (\citealt{tri95}; \citealt{kob99}).
For iron, the simulated Fe gradients are steeper than the observation, even if one takes into account the much larger observational errors of Fe$_1$ than Mg$_2$.
This is because in the simulation the star formation does not terminate completely in the galaxy center (see \S \ref{sec:discussion}).

From the observed metallicity gradients, we can estimate the fraction non-major merger and major merger galaxies.
Using the Mg$_2$ observation, we derive the numbers of nearby elliptical galaxies with steep gradients 
of $\Delta\log Z/\Delta\log r \le -0.3$ and flat gradients of $> -0.3$ 
as 25 and 21, respectively. 
If we take alternative threshold value of $\Delta\log Z/\Delta\log r = -0.25$, the numbers become 30 and 16, respectively. 
Therefore, the fraction of non-major merger galaxies is half or two third.
Even if we take account of some problems involved in our initial condition 
(see \S \ref{sec:discussion} for the detail), 
there exist non-major merger galaxies and major merger galaxies half and half.
The observed variation in the metallicity gradients cannot be explained by either 
the {\it monolithic collapse} only or the {\it major merger} only.
It is well reproduced in the present model where
both formation processes arise under the CDM scheme.

\subsection{EVOLUTION OF GRADIENTS VIA MERGERS}
\label{sec:evozg}

The spatial distribution of metallicity is already shown in the last lines of Figures \ref{fig:evomapD} ([1] the monolithic case) and \ref{fig:evomapM} ([4] the major merger case).
The metallicity is enhanced in the high density region, and radial metallicity gradients appear in protogalaxies at $z \sim 5$.
After the major merger at $z \simeq 2.0$, many metal-rich stars move to the outer region of the galaxy.
In the monolithic case, metal-rich region is concentrated, while metal-rich region extends over $\pm 10$ kpc in the major merger case.
The present metallicity gradients are $\Delta\log Z/Z_\odot/\Delta\log r = -0.46$ and $-0.19$, respectively.

\begin{figure}
\begin{center}
\includegraphics[height=8cm,angle=-90]{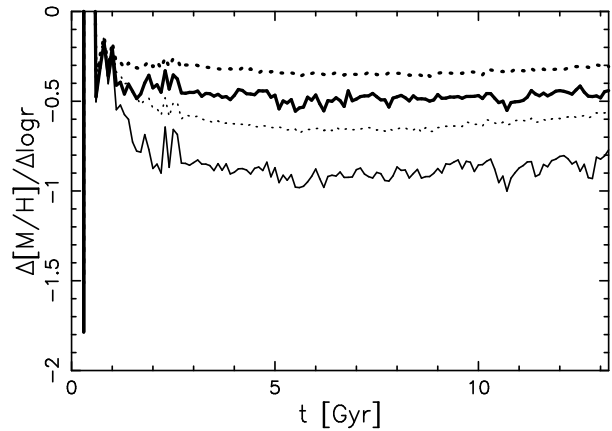}
\caption{\label{fig:evocgD}
The evolution of metallicity gradients for the monolithic collapse case.
The thick and thin lines are for the oxygen and iron abundances, respectively. 
The solid and dotted lines are the gradients weighted by V-luminosities and masses, respectively.
}
\end{center}
\end{figure}

\begin{figure}
\begin{center}
\includegraphics[height=8cm,angle=-90]{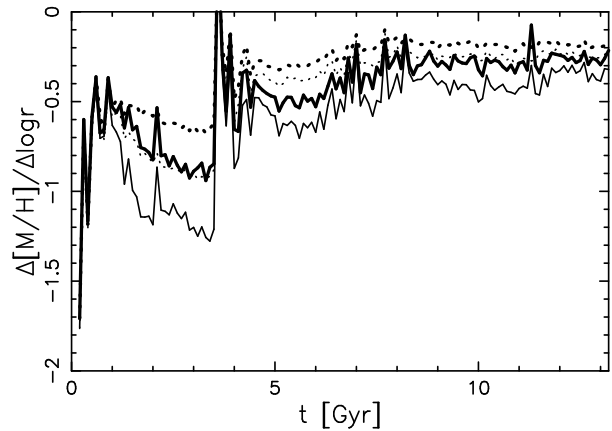}
\caption{\label{fig:evocgM}
The same as Figure \ref{fig:evocgD}, but for the major merger case.
Mergers takes place at $t \simeq 3.7, 5.1$, and $6.9$ Gyr.
}
\end{center}
\end{figure}

Figures \ref{fig:evocgD} and \ref{fig:evocgM} show the time evolutions of metallicity gradients 
for [1] the monolithic case and [5] the multiple major merger case, respectively.
The thick and thin lines are for the oxygen and iron abundances, respectively.
The solid and dotted lines indicate the gradients weighted by V-luminosity and mass, respectively.
The gradients are underestimated if the metallicity is weighted by mass.

Figure \ref{fig:evocgD} shows the gradient evolution of ID H3/782389913.
This galaxy forms through the assembly of 3 galaxies with the stellar masses of 
$8 \times 10^9 M_\odot$, $4 \times 10^9 M_\odot$, and $4 \times 10^9 M_\odot$, and thus the 
initial gradient is not so steep, $\Delta{\rm [Fe/H]}/\Delta\log r \simeq -0.9$.
This assembly has finished by $z \simeq 3.7$ ($t \simeq 1.2$ Gyr).
Then the galaxy evolves quietly, and the metallicity gradient is kept nearly constant. 
At $z \simeq 0.1$ ($t \simeq 10$ Gyr), a small galaxy with $\sim 10^8 M_\odot$ accretes. 
The small star formation is induced, which changes the gradient by $0.1$ dex temporarily.

Figure \ref{fig:evocgM} shows the gradient evolution of ID H3/390367807.
This galaxy undergoes the major mergers three times at $z \simeq 1.3, 0.9$, and $0.5$ 
($t \simeq 3.7, 5.1$, and $6.9$ Gyr). The mass ratios are $f=0.47, 0.16$, and $0.24$, respectively. 
The initial star burst produces the metallicity gradient of $\Delta{\rm [Fe/H]}/\Delta\log r \sim -1.2$. 
The major merger event is so violent that the gradient decreases by $0.7$ dex to be $\Delta{\rm [Fe/H]}/\Delta\log r \simeq -0.5$.
The second and third merger events are not so active, and the gradient becomes 
$\Delta{\rm [Fe/H]}/\Delta\log r \simeq -0.4$ and $-0.3$ at the end of the merger event, respectively.

The initial gradient is determined during the initial star burst at $z\gtsim3$.
The gradient is steeper in the case of quiescent gas accretion, and is shallower in the case of 
violent assembly of subgalaxies.
As a result, the initial gradients span from $\Delta{\rm [Fe/H]}/\Delta\log r \simeq -1.5$ to $-1.0$.
Since the gradients evolve a lot after this time, the initial gradients cannot be inferred from 
the present gradients.

\begin{figure*}
\begin{center}
\includegraphics[width=16cm]{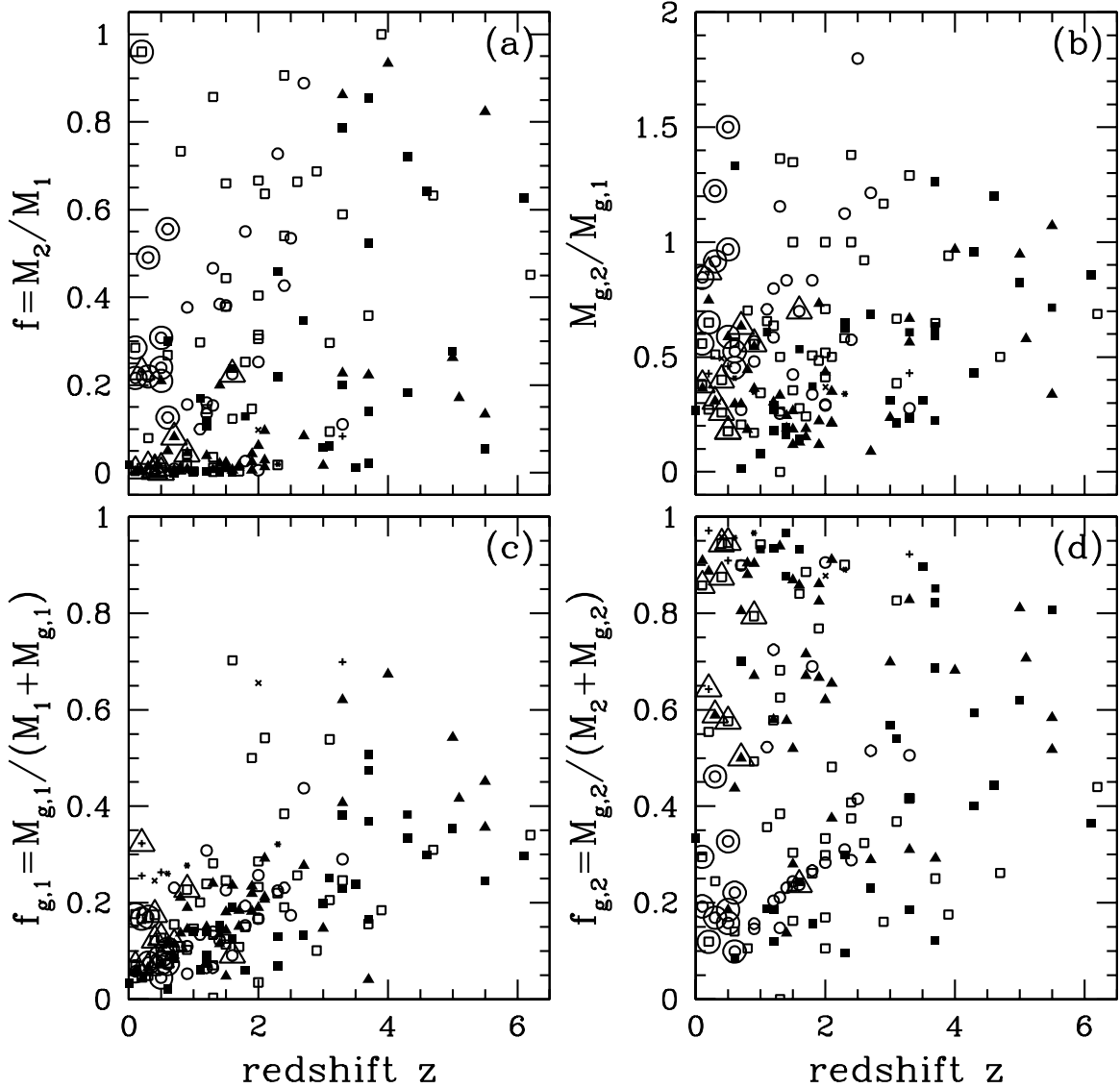}
\caption{\label{fig:mergerz}
(a) Mass ratios $f\equiv M_2/M_1$ ($M_1\ge M_2$), (b) ratios of gas mass 
$M_{\rm g,2}/M_{\rm g,1}$, and (c,d) gas fractions ($f_{\rm g,1}$ and $f_{\rm g,2}$) 
of the primary and secondary galaxies against redshifts $z$.
For the symbols, see the caption of Figure \ref{fig:grad}.
The points surrounded by large circles/triangles indicate that strong/moderate star formation is induced.
}
\end{center}
\end{figure*}

\begin{figure*}
\begin{center}
\includegraphics[width=16cm]{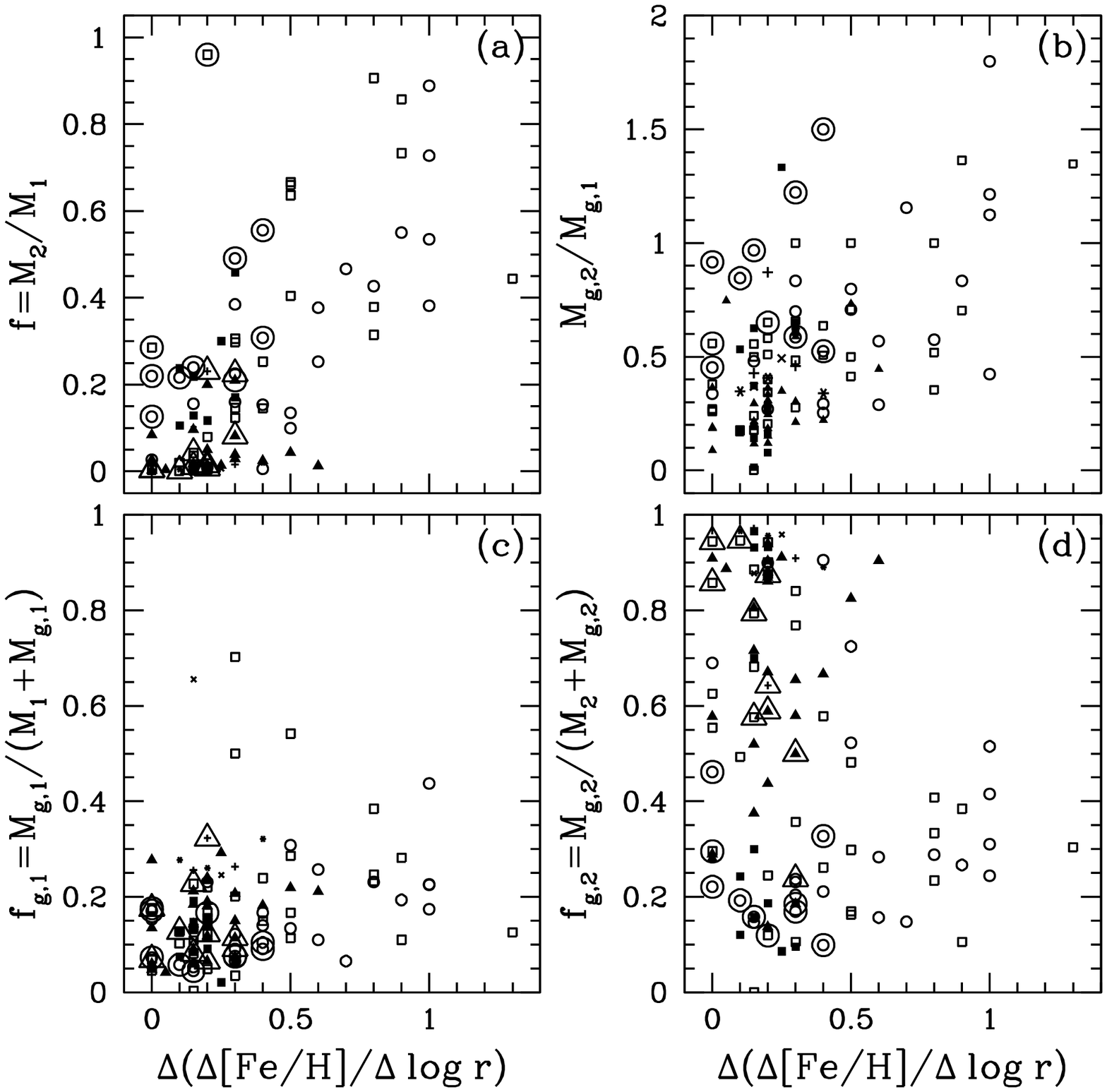}
\caption{\label{fig:mergerg}
The gradient change $\Delta(\Delta{\rm [Fe/H]}/\Delta\log r)$ before and after the merging event at $z<3$ 
against (a) mass ratios $f$, (b) ratios of gas mass $M_{\rm g,2}/M_{\rm g,1}$, and 
(c,d) gas fractions ($f_{\rm g,1}$ and $f_{\rm g,2}$).
For the symbols, see the caption of Figure \ref{fig:grad}.
The points surrounded by large circles/triangles indicate that strong/moderate star formation is induced.
}
\end{center}
\end{figure*}

The metallicity gradients become definitely shallower when galaxies merge.
However, the formation and destruction of gradients via mergers are complicated. 
To find out some rules, we examine the physical conditions at the 151 merging events 
that occur in our simulations of giant galaxies.
Figure \ref{fig:mergerz} shows mass ratios $f\equiv M_2/M_1$ ($M_1\ge M_2$) (a), 
ratios of gas mass $M_{\rm g,2}/M_{\rm g,1}$ (b), gas fractions of primary galaxies $f_{\rm g,1}$ (c), 
and those of secondary galaxies $f_{\rm g,2}$ (d) against redshifts $z$.
The averages of $f$ and $f_{\rm g,1}$ in each redshift bin 
decrease toward $z=0$ because the stellar masses of primary galaxies increase.
The average of $M_{\rm g,2}/M_{\rm g,1}$ is larger than that of $f$, and is constant of $\sim 0.5$. 
$f_{\rm g,2}$ is larger than $f_{\rm g,1}$, and spans from $0$ to $1$.
These are because star formation strongly depends on gas density and takes place slowly in the less-massive secondary galaxies.
At $z<1$, some major mergers ($f \gtsim 0.2$) have $M_{\rm g,2}/M_{\rm g,1} \gtsim 0.5$, 
which means that the secondary galaxies have comparable mass and gas content to the primaries.
These induce strong star formation (surrounded by circles). 
Relatively weaker star formation (surrounded by triangles) is induced by mergers with 
$f_{\rm g,2} \gtsim 0.5$, which means that the stellar mass of the secondary is small and 
the secondary galaxy behave like a gas cloud.

Figure \ref{fig:mergerg} shows the change of gradient, $\Delta(\Delta{\rm [Fe/H]}/\Delta\log r)$, caused by a merging event at $z<3$ against various physical conditions: $f$ (a), $M_{\rm g,2}/M_{\rm g,1}$ (b), $f_{\rm g,1}$ (c), and $f_{\rm g,2}$ (d).
The gradient change strongly depends on the mass ratio $f$. 
The mergers with $f>0.2$ decrease the gradients at least by $0.5$ dex. 
This is the reason why we have defined major mergers as those with $f>0.2$.
A major merger changes the orbits of stars. The stars that are in the center 
and have high metallicities are able to move to the outer region of the galaxy.
However, if the secondary galaxy contains as much gas as the primary galaxy 
(i.e., $M_{\rm g,2}/M_{\rm g,1} \gtsim 0.5$) and strong star formation is induced 
(surrounded by circles), then the gradient change stays smaller than $0.5$ dex. 
This is because such star formation takes place after the gas falls into the center of 
the primary galaxy and increases the metallicity at the center, resulting in a small gradient change.
Sometimes merging of a gas rich galaxy ($f_{\rm g,2} \gtsim 0.5$) also induces moderate star formation 
(surrounded by triangles). In this case, star formation takes place in the outer 
region of the galaxy, and the gradient change can be as large as $\sim 0.5$ dex, even if $f\sim 0$.
In some cases without a merging event, a similar star formation is induced by late 
gas accretion, and the metallicity gradient gradually becomes shallower.

\begin{figure}
\begin{center}
\includegraphics[width=7cm]{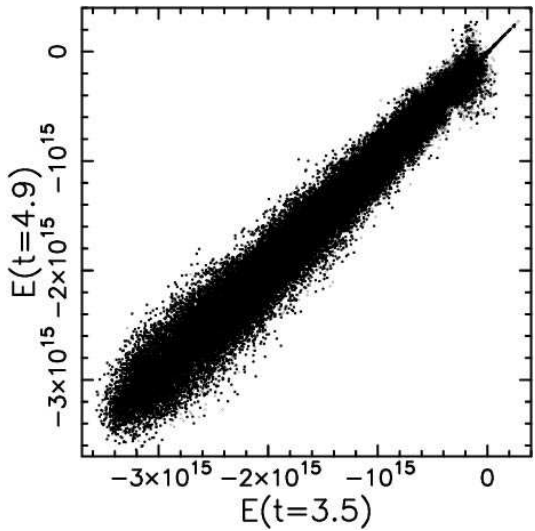}
\caption{\label{fig:enenD}
The energies $E \equiv v^2/2 + \Phi$ of particles in the non-merger galaxy.
}
\end{center}
\end{figure}

\begin{figure}
\begin{center}
\includegraphics[width=7cm]{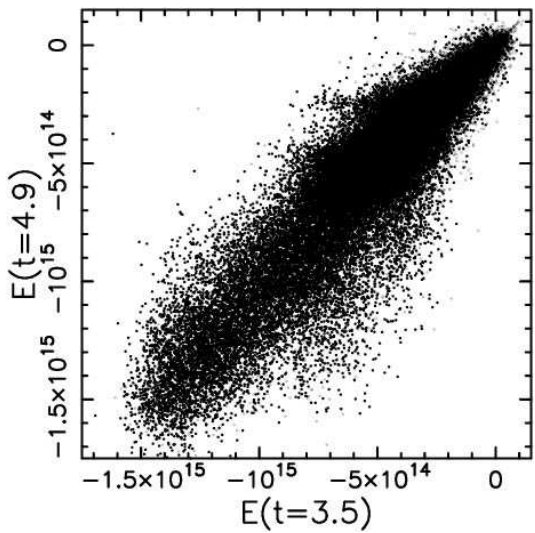}
\caption{\label{fig:enenM}
The energies of particles before and after the major merger at $\sim 3.7$ Gyr.
}
\end{center}
\end{figure}

By showing the similar figure in \citet{vanalb82},
we argue that dynamical information on the orbits of N-body particles is not fully wiped out by a merger, but fairly lost to change the metallicity gradient. 
Figures \ref{fig:enenD} and \ref{fig:enenM} show the energies $E \equiv v^2/2 + \Phi$ of particles before and 
after a merging event.
Large $E$ means that a particle has an extended orbit.
Figure \ref{fig:enenM} shows the energies at 3.5 and 4.9 Gyr for the first major merger at $\simeq 3.7$ Gyr of 
the galaxy ID H3/390367807 that undergoes a triple major merger (see Figure \ref{fig:evocgM} for the gradient evolution). 
Figure \ref{fig:enenD} shows the same but for the non-merger galaxy ID H3/782389913. Clearly, the dispersion for the merger galaxy is much larger than the non-merger galaxy.

In summary, whether the merging event changes the metallicity gradient is mainly influenced by two factors; 1) the mass 
ratio of the merging galaxies $f$ and 2) the induced star formation.
The evolution of the gradients via merging events is based on the following three processes: 
i) Destruction by mergers to an extent dependent on $f$.
ii) Regeneration due to the central star formation induced at a rate dependent on $M_{\rm g,2}/M_{\rm g,1}$.
iii) Passive evolution as star formation is induced in the outer regions at a rate dependent on $f_{\rm g,2}$.
We note that the gradient change is underestimated if the metallicity is weighted not by luminosity as the observation, but by mass as in previous simulations.

\section{DISCUSSION}
\label{sec:discussion}

The global properties of elliptical galaxies depend mainly on their masses, 
while their metallicity gradients are much affected by their merging history.
Merging histories can thus, in principle, be inferred from the observed 
metallicity gradients of present-day galaxies.
The dispersion in metallicity gradients for galaxies with similar merging histories is not small (Fig. \ref{fig:gh}), and so it may be 
difficult to determine the merging history of an individual galaxy from its metallicity gradient. 
However, it should be possible to estimate the fractions of non-merger and merger galaxies 
by using the statistics of metallicity gradients. For example, if the fractions are estimated for field and cluster galaxies, 
it should provide information about environmental effects on galaxy formation.
Even allowing for the uncertainties in available observations of nearby galaxies, our predicted difference between gradients of non-merger 
and merger galaxies is large enough to be detected.

\begin{figure}
\begin{center}
\includegraphics[width=6cm]{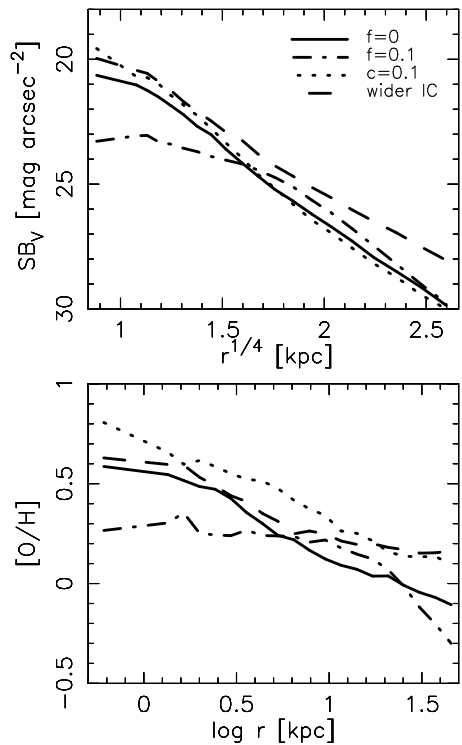}
\caption{\label{fig:param}
Surface brightness profiles (upper panel) and metallicity gradients (lower panel) with kinetic feedback (dash-dotted line, $f_{\rm kin}=0.1$), with longer timescale of the star formation (dotted line, $c=0.1$), and wider region of initial condition ($2.5$ Mpc).
}
\end{center}
\end{figure}

We should note that there are two problems in the simulated galaxies, which cannot be improved by changing 
parameters in the present model.
One is that galactic winds do not occur in large galaxies in the simulations, and star formation never 
terminates completely. The colors of our galaxies thus tend to be bluer than the 
observation, and the B-V color distribution extends to 0.4 mag bluer.
In our galaxy centers, 
the color gradients tend to have the opposite slope to the metallicity gradients, because the stellar populations there 
are young. 
Such late star formation takes place in the central $1$ kpc, and this region is carefully excluded 
when estimating metallicity gradients in this paper.
This problem arises from our SPH method and our feedback scheme.
If we include the kinetic feedback with $f_{\rm kin}>0$, surface brightness decreases at the center, and metal-rich gas blow out. 
However, these results in too large effective radii and too shallow metallicity gradients to meet the observation.
Figure \ref{fig:param} shows the surface brightness profiles and metallicity gradients with (dash-dotted line) and without (solid line) $f_{\rm kin}$ using the same initial condition.
With $f_{\rm kin}=0.1$, the surface brightness decreases by $\sim 2$ mag at the center, which results in $r_{\rm e}=8.1$ kpc ($r_{\rm e}=4.3$ kpc for $f_{\rm kin}=0$). The metallicity gradient is almost flat within $r_{\rm e}$.
To describe galactic winds with the kinetic feedback, star formation scheme should be modified simultaneously.

We note that our predictions for oxygen abundance are not changed so much, whereas the iron abundance 
should decrease especially at the galaxy center. 
In Figure \ref{fig:grad}, there appear to be a relation between the iron abundance and iron gradients.
This is because late star formation at the galaxy center increases 
the iron abundance and steepens iron gradients at the same time.
Such a relation can disappear if the galaxy model is improved.
In Figure \ref{fig:gh}, the mean value of simulated iron gradients is steeper than observed.
This is also caused by the same reason.
If the galaxy model is improved so as to blow a galactic wind, 
iron gradients for non-merger galaxies will become the similar to their oxygen gradients.
However, iron gradients for merger galaxies will be still steeper than oxygen gradients. 
The difference of iron and oxygen gradients, i.e., [O/Fe] gradients, may be the best indicator of the formation history of the galaxy.
To discuss [O/Fe], observational data of high-enough S/N is required, and the spectral population synthesis models should be prepared for various [O/Fe].

The other problem is that simulated galaxies are more extended, and thus there is an offset in the radius-magnitude relation.
This problem is not in magnitude, because the Faber-Jackson relation can be reproduced, and the galaxy with given mass has reasonable luminosity.
While the stellar mass-to-light ratio is consistent with the observation as $M/L_{\rm B} \sim 5-8$, the total mass-to-light ratio is as large as $(M_{\rm DM}+M_{\rm baryon})/L \sim 10-100$ (within a sphere of $2 r_{\rm e}$), which may be too large.
The baryon fraction increases to $\sim 0.5$ at the center ($r<1$ kpc) because of the stellar concentration, but is as small as $\sim 0.2$ in $r<2 r_e$.
Baryon does not fully dominate even at the galaxy center in our simulated galaxies, and
the stellar concentration seems to be not enough compared with the dark matter concentration.
These might be due to cosmological parameters, dynamical friction, star formation parameter, and the limited field size of initial condition.
As well known with the zero-point offset in the Tully-Fisher relation (\citealt{nav00}), such high dark matter concentration should arise from the standard CDM cosmology that is adopted in this paper, and should decrease to some extent with the $\lambda$-CDM cosmology.
However, \citet{nav00} argued that the zero-point offset remains even with the $\lambda$-CDM cosmology, which requires substantial revision to the CDM scenario such as $\sigma_8$, a tilt in the primordial power spectrum, or hot dark matter.
The dynamical friction may be effective because the mass of a dark matter particle is ten times larger 
than the gas and stellar particles (but see \citealt{ste97}).
Too large effective radius suggests that 
the star formation takes place too early before the gas accretes 
towards the center.
This can be solved by changing the star formation timescale, i.e., reducing the star formation parameter $c$, although colors become bluer.
As shown in Figure \ref{fig:param}, with $c=0.1$ (dotted line), the surface brightness increase by 1 mag at the center, which results in smaller effective radius as $r_{\rm e}=3.2$ kpc.
The metallicity increases because of longer timescale of star formation, but the gradient dose not change so much.
As mentioned in \S \ref{sec:fit}, the surface brightness profiles of our simulated galaxies well follow the de-Vaucouleurs law, but is smeared out at the center, and the central surface brightness ($r<1$ kpc) is slightly smaller than the de-Vaucouleurs fit (see Fig.\ref{fig:param}).
Although the resolution is not enough to discuss, this central stellar concentration may be smaller than other simulations (e.g., \citealt{spr03}), can be as large with $c=0.1$, and should be affected by the difference in the star formation and feedback schemes.

The initial condition is most important.
The edge of the simulated fields actually falls into a galaxy at $z \sim 2-4$ depending on the position of the galaxy in the field.
The galaxy mass $M_{200}$ (the mass with higher density than $200\rho_{\rm crit}$) stops increasing at $z \sim 1.5-3$.
This means that there is an artificial cutoff of mass accretion around this redshift, which results in the cutoff of star formation rate in Figure \ref{fig:sfr}.
In a wider simulation with a radius of $\sim 2.5$ Mpc (dashed-line in Fig.\ref{fig:param}) instead of $\sim 1.5$ Mpc in the series of our simulation, the star formation continues longer, and thus colors tend to be too blue.
The total luminosity becomes $\sim 2$ times larger, but still $\sim 3$ times smaller than the observation.
With much wider simulation, the luminosity difference could be improved, but the color inconsistency become much larger, and such galaxy is no more elliptical.
The star formation and feedback schemes need to be modified accordingly, such as some recent works have tried (e.g., \citealt{spr03}).
In this paper, the mass accretion and star formation are truncated artificially by the initial condition.
However, in observed ellipticals, star formation should be truncated at $z \sim 2$ by some process; tidal stripping, effects of active galactic nuclei, and so on.
This epoch may correspond to the epoch when the galaxy falls into a cluster, because the galaxy moves too fast to arise mass accretion.

If we include the contribution of the outside of the simulated field, late accretion and late merging 
events should increase. Both make the metallicity gradient shallow.
As noted in \S \ref{sec:init}, fields with larger spin parameter $\lambda$ are not included in our simulation, and
an elliptical can form through major mergers of disk galaxies in some of such fields.
However, as shown in Figure \ref{fig:gh}, the predicted metallicity gradients are already in good agreement with the observation, and the increase of merger galaxies conflicts with the observation.
If star formation and feedback schemes improved, evolution of metallicity gradients may also be affected at some extent.
We should note, however, that the following conclusions will not change; major merger make the metallicity gradients shallow, and there should be non-merger ellipticals to explain the observed steep gradients.

\section{CONCLUSIONS}
\label{sec:conclusion}

We study the formation and chemodynamical evolution of galaxies with our GRAPE-SPH chemodynamical model 
that includes various physical processes associated with the formation of stellar systems; radiative cooling, 
star formation, feedback from SNe II, SNe Ia, and SWs, and chemical enrichment.
We simulate 72 slowly-rotating spherical fields (spin parameter $\lambda \sim 0.02$), and obtain 124 galaxies 
(78 ellipticals and 46 dwarfs) from the CDM initial fluctuation.
All simulated galaxies have the de Vaucouleurs' surface brightness profiles, 
and are therefore elliptical galaxies.
Most stars in ellipticals form during the initial star burst at $z \gtsim 2$, 
while dwarfs undergo relatively continuous star formation.
In our scenario, galaxies form through the successive merging of sub-galaxies. 
The merging history is various and the difference is seeded in the initial conditions.
In some cases, galaxies form through the assembly of gas rich small galaxies, and the process looks like a {\it monolithic collapse}. 
In other cases, the evolved galaxies undergo {\it major merger} of galaxies.
Major mergers are defined as those with the mass ratios of the primary and secondary galaxies being $f \gtsim 0.2$.

We examine the physical conditions during 151 merging events that occur in our simulation.
Whether the merging event changes the metallicity gradient is mainly influenced by two factors; the mass ratio of the merging galaxies $f$ and the induced star formation.
The basic processes of the formation and evolution of the gradients are summarized below:
\begin{itemize}
\item Formation of initial gradients ---
The initial gradient is determined from the initial star burst at $z\gtsim3$.
The gradient is steeper in the case of quiescent gas accretion, and is shallower in the 
case of violent assembly of subgalaxies. 
As a result, the initial gradients span from 
$\Delta{\rm [Fe/H]}/\Delta\log r = -1.5$ to $-1.0$.
\item Destruction by mergers ---
The major merger changes the orbits of stars. 
The metal-rich stars at the center are able to move to the outer region of the galaxy.
The gradient change is determined mainly from the mass ratio of merging galaxies $f$. 
With larger $f$, the gradients become shallower.
If the mass ratio of merging galaxies is larger than $f\sim 0.2$, the gradient change is larger than $\sim 0.5$ dex.
\item Regeneration due to the induced star formation ---
If the ratio of gas mass is as large as $M_{\rm g,2}/M_{\rm g,1}\gtsim 0.5$, strong star formation is 
induced at the center of the primary galaxy, and the gradient change is smaller than $\sim 0.5$ dex.
\item Passive evolution ---
If the gas fraction of the secondary galaxy is larger than $f_{\rm g,2}\sim 0.5$, moderate star formation is 
induced in the outer region of the primary galaxy, and the gradient change becomes as large as $\sim 0.5$ dex, even if $f \sim 0$.
In some case without merging event, if the similar star formation is induced by the late gas accretion, 
the metallicity gradient gradually becomes shallower.
\end{itemize}

We succeed in reproducing the observations of metallicity gradients
and finding the origin of the variety of internal structures.
From the distribution functions of the gradients for different merging histories,
we discuss the origin of elliptical galaxies.
\begin{itemize}
\item
The average metallicity gradient is $\Delta\log Z/\log r \simeq -0.3$ and 
the dispersion is $\pm 0.2$, which are both 
consistent with observations of Mg$_2$ gradients.
\item
No correlation is produced between gradients and masses.
The metallicity gradients do not depend on the galaxy mass, and the variety of the gradients stems from 
the difference in the merging histories;
galaxies that form monolithically have steeper gradients, while galaxies 
that undergo major mergers have shallower gradients.
\item
The metallicity gradient distributions for [A] non-major merger ([1]-[3]) and 
[B] major merger galaxies ([4] and [5]) are quite different.
The typical gradients for non-major merger and major merger galaxies are 
$\Delta\log Z/\Delta\log r \sim -0.3$ and $-0.2$, respectively.
Simulated galaxies with the gradients steeper than $-0.35$ are all non-major merger galaxies.
\end{itemize}
 
The global properties of elliptical galaxies depend mainly on their masses, 
while their metallicity gradients are much affected by their merging history.
A major merger makes the gradient shallower. 
Therefore, merging histories can be inferred from the observed metallicity gradients of present-day galaxies.
Available observations for nearby galaxies suggest that there exist non-major merger galaxies 
and major merger galaxies half and half.  
The observed variation in the metallicity gradients cannot be explained by either 
{\it monolithic collapse} or by {\it major merger} alone.
Instead, it is well reproduced in the present model in which
both formation processes arise under the CDM scheme.

\section*{Acknowledgments}

This paper is a part of the Ph.D. thesis of C. Kobayashi in the Astronomy Department of the University of Tokyo.
I would like to thank the supervisor, K. Nomoto and the advisor, N. Arimoto.
I am grateful to
N. Nakasato, J. Makino, M. Mori, T. Kodama, N. Tamura, K. Ohta, V. Springel, F. van den Bosch, and A. Renzini for fruitful discussion, and S.D.M. White  for detailed suggestion.
I also thank to the Japan Society for Promotion 
of Science for a financial support,
and to National Observatory of Japan for the GRAPE-5 MUV system.

\bsp

\label{lastpage}


\begin{thebibliography}{}

\bibitem[\protect\citeauthoryear{Abbott}{1982}]{abb82}
Abbott, D. C. 1982, \apj, 263, 723

\bibitem[\protect\citeauthoryear{Arimoto \& Yoshii}{1987}]{ari87} 
Arimoto, N., \& Yoshii, Y. 1987, \aap, 173, 23

\bibitem[\protect\citeauthoryear{Bacon et al.}{2001}]{bac01}
Bacon, R., et al. 2001, \mnras, 326, 23

\bibitem[\protect\citeauthoryear{Balsara}{1995}]{bal95}
Balsara, D. S. 1995, J. Chem. Phys., 121, 357

\bibitem[\protect\citeauthoryear{Bargar et al.}{1999}]{bar99}
Bargar, A. J., Cowie, L. L., Trentham, N., Fulton, E., Hu, E. M., Songaila, A., \& Hall, D. 1999, \aj, 117, 102

\bibitem[\protect\citeauthoryear{Barnes}{1988}]{bar88}
Barnes, J. E. 1988, \apj, 331, 699

\bibitem[\protect\citeauthoryear{Barnes}{1996}]{bar96}
Barnes, J. E. 1996, in IAU Symp. 171, 
New Light on Galaxy Evolution, p.191

\bibitem[\protect\citeauthoryear{Baugh, Cole \& Frenk}{1996}]{bau96}
Baugh, C. M., Cole, S., \& Frenk, C. S. 1996, \mnras, 283, 1361

\bibitem[\protect\citeauthoryear{Bender \& Surma}{1992}]{ben92}
Bender, R., \& Surma, P. 1992, \aap, 258, 250

\bibitem[\protect\citeauthoryear{Benson, Ellis \& Menanteau}{2002}]{ben02}
Benson, A. J., Ellis, R. S., \& Menanteau, F. 2002, \mnras, 336, 564

\bibitem[\protect\citeauthoryear{Benz et al.}{1990}]{ben90}
Benz, W., Bowers, R., Cameron, A., Press, W. 1990, \apj, 348, 647

\bibitem[\protect\citeauthoryear{Bertschinger}{1987}]{ber87}
Bertschinger, E. 1987, \apj, 323, L103

\bibitem[\protect\citeauthoryear{Bertschinger}{1995}]{ber95}
Bertschinger, E. 1995, http://arcturus.mit.edu/cosmics/

\bibitem[\protect\citeauthoryear{Bower, Lucey \& Ellis}{1992}]{bow92} 
Bower, R. G., Lucey, J. R., \& Ellis, R. S. 1992, \mnras, 254, 601

\bibitem[\protect\citeauthoryear{Blinnikov et al.}{2000}]{bli00}
Blinnikov, S., Lundqvist, P., Bartunov, O., Nomoto, K., \& Iwamoto, K. 2000, \apj, 532, 1132

\bibitem[\protect\citeauthoryear{Brinchmann \& Ellis}{2000}]{bri00}
Brinchmann, J., \& Ellis, R. S. 2000, \apj, 536, L77

\bibitem[\protect\citeauthoryear{Carlberg}{1984}]{car84} 
Carlberg, R. G. 1984, \apj, 286, 403

\bibitem[\protect\citeauthoryear{Carollo, Danziger \& Buson}{1993}]{car93} 
Carollo, C. M., Danziger, I. J., \& Buson, L. 1993, \mnras, 265, 553

\bibitem[\protect\citeauthoryear{Carraro, Lia \& Chiosi}{1998}]{car98}
Carraro, G. Lia, C., \& Chiosi, C. 1998, \mnras, 297, 1021

\bibitem[\protect\citeauthoryear{Cimatti et al.}{2002}]{cim02}
Cimatti, A. et al. 2002, \aap, 391, L1

\bibitem[\protect\citeauthoryear{Ciotti et al.}{1991}]{cio91} 
Ciotti, L., D'Eecole, A., Pellegrini, S., \& Renzini, A. 1991, \apj, 376, 380

\bibitem[\protect\citeauthoryear{Cole et al.}{1994}]{col94} 
Cole, S., Arag\'on-Salamanca, A., Frenk, C. S., Navarro, J. F., \& Zepf, S. E. 
1994, \mnras, 271, 781

\bibitem[\protect\citeauthoryear{Cole et al.}{2000}]{col00} 
Cole, S., Lacey, C. G., Baugh, C. M., \& Frenk, C. S. 2000, \mnras, 319, 168

\bibitem[\protect\citeauthoryear{Daddi, Cimatti \& Renzini}{2000}]{dad00}
Daddi, E., Cimatti, A., \& Renzini, A. 2000, \aap, 362, L45

\bibitem[\protect\citeauthoryear{David, Forman \& Jones}{1990}]{dav90} 
David, L. P., Forman, W., \& Jones, C. 1990, \apj, 459, 29

\bibitem[\protect\citeauthoryear{Davies et al.}{1987}]{dav87} 
Davies, R. L., Burstein, D., Dressler, A., Faber, S. M., Lynden-Bell, D., 
Terlevich, R. J., \& Wegner, G. 1987, \apjs, 64, 581

\bibitem[\protect\citeauthoryear{Davies, Sadler \& Peletier}{1993}]{dav93} 
Davies, R. L., Sadler, E. M., \& Peletier, R. F. 1993, \mnras, 262, 650

\bibitem[\protect\citeauthoryear{de Vaucouleurs}{1961}]{dev61}
de Vaucouleurs, G. 1961, \apjs, 5, 233

\bibitem[\protect\citeauthoryear{Dickinson}{1996}]{dic96} 
Dickinson, M. 1996, in ASP Conference Series, 86, Fresh Views of 
Elliptical Galaxies, eds., A.Buzzoni, A.Renzini, \& A.Serrano, p.283

\bibitem[\protect\citeauthoryear{Dickinson et al.}{2003}]{dic03} 
Dickinson, M., Papovich, C., Ferguson, H. C., \& Budav\'ari, T. 2003. astro-ph/0212242

\bibitem[\protect\citeauthoryear{Djorgovski \& Davis}{1987}]{djo87} 
Djorgovski, S., \& Davis, M., 1987, \apj, 313, 59

\bibitem[\protect\citeauthoryear{Dressler}{1980}]{dre80} 
Dressler, A. 1980, \apj, 236, 351

\bibitem[\protect\citeauthoryear{Dressler et al.}{1987}]{dre87} 
Dressler, A., Lynden-Bell, D., Burstein, D., Davies, R. L., Faber, S. M., Terlevich, R. J., \& Wegner, G. 1987, \apj, 313, 42

\bibitem[\protect\citeauthoryear{Dressler et al.}{1997}]{dre97} 
Dressler, A., Oemler, A., Couch, W. J., Smail, I., Ellis, R. S., Barger, A., Butcher, H.,  Poggianti, B. M., \& Sharples, R. M. 1997, \apj, 490, 577

\bibitem[\protect\citeauthoryear{Drory et al.}{2001}]{dro01}
Drory, N., Bender, R., Snigula, J., Feulner, G., Hopp, U., maraston, C., Hill, G. J., \& Menndes de Oliveira, C. 2001, \apj, 562, L111

\bibitem[\protect\citeauthoryear{Efstathiou et al.}{1985}]{efs85}
Efstathiou, G., Davis, M., Frenk, C. S., \& White, S. D. M. 1985, \apjs, 57. 241

\bibitem[\protect\citeauthoryear{Evrard}{1988}]{evr88}
Evrard, A. E. 1988, \mnras, 235, 911

\bibitem[\protect\citeauthoryear{Faber}{1973}]{fab73} 
Faber, S. M. 1973, \apj, 179, 731

\bibitem[\protect\citeauthoryear{Faber}{1977}]{fab77}
Faber, S. M. 1977, in The Evolution of Galaxies and Stellar Populations, ed. B.T. Tinsley, \& R. B. Larson (New Heaven:Yale Univ. Press), p.157

\bibitem[\protect\citeauthoryear{Fontana et al.}{1999}]{fon99}
Fontana, A., Menci, N., D'Odorico, S., Giallongo, E., Poli, F., Cristiani, S., Moorwood, A., \& Saracco, P. 1999, \mnras, 310, L27

\bibitem[\protect\citeauthoryear{Franceschini et al.}{1998}]{fra98}
Franceschini, A., Silva, L., Fasano, G., Granato, G. L., Bressan, A., 
Arnouts, S., \& Danese, L. 1998, \apj, 506, 600

\bibitem[\protect\citeauthoryear{Gingold \& Monaghan}{1977}]{gin77}
Gingold, R. A., \& Monaghan, J. J. 1977, MNRAS, 181, 375

\bibitem[\protect\citeauthoryear{Greggio \& Renzini}{1983}]{gre83} 
Greggio, L., \& Renzini, A. 1983, \aap, 118, 217

\bibitem[\protect\citeauthoryear{Gonzalez \& Gorgas}{1996}]{gon96} 
Gonz\'alez, J. J., \& Gorgas, J. 1996, in ASP Conference Series, 86, 
Fresh Views of Elliptical Galaxies, eds., A.Buzzoni, A.Renzini, \& 
A.Serrano, p.225

\bibitem[\protect\citeauthoryear{Gorgas, Efstathiou \& Arag\'on-Salamanca}{1990}]{gor90} 
Gorgas, J., Efstathiou, G., \& Arag\'on-Salamanca, A. 1990, \mnras, 245, 217

\bibitem[\protect\citeauthoryear{Gott}{1973}]{got73}
Gott, J. R. 1973, \apj, 186, 481

\bibitem[\protect\citeauthoryear{Gott}{1975}]{got75}
Gott, J. R. 1975, \apj, 201, 296

\bibitem[\protect\citeauthoryear{Habe \& Ohta}{1992}]{hab92}
Habe, A., \& Ohta, K. 1992, \pasj, 44, 203

\bibitem[\protect\citeauthoryear{Hernquist \& Katz}{1989}]{her89}
Hernquist, L., \& Katz, N. 1989, \apjs, 70, 419

\bibitem[\protect\citeauthoryear{Im et al.}{2002}]{im02}
Im, M. et al. 2002, \apj, 571, 136

\bibitem[\protect\citeauthoryear{Katz}{1992}]{kat92}
Katz, N. 1992, \apj, 391, 502

\bibitem[\protect\citeauthoryear{Katz, Weinberg \& Hernquist}{1996}]{kat96}
Katz, N., Weinberg, D. H., \& Hernquist, L. 1996, \apjs, 105, 19 

\bibitem[\protect\citeauthoryear{Kauffmann}{1996}]{kau96}
Kauffmann, G. 1996, \mnras, 281, 487 

\bibitem[\protect\citeauthoryear{Kauffmann \& Charlot}{1998}]{kau98} 
Kauffmann, G., \& Charlot, S. 1998, \mnras, 294, 705

\bibitem[\protect\citeauthoryear{Kauffmann, White \& Guiderdoni}{1993}]{kau93} 
Kauffmann, G., White, S. D. M., \& Guiderdoni, B. 1993, \mnras, 264, 201

\bibitem[\protect\citeauthoryear{Kawata \& Gibson}{2003}]{kaw03}
Kawata, D., \& Gibson, B. K. 2003, astro-ph/0212401

\bibitem[\protect\citeauthoryear{Kelson et al.}{1997}]{kel97} 
Kelson, D. D., van Dokkum, P. G., Franx, M., Illingworth, G. D., \& Fabricant, D. 1997, \apj, 478, L13

\bibitem[\protect\citeauthoryear{Kennicutt}{1983}]{ken83}
Kennicutt, Jr. R. C. 1983, \apj, 272, 54

\bibitem[\protect\citeauthoryear{Kennicutt}{1989}]{ken89}
Kennicutt, Jr. R. C. 1989, \apj, 344, 685

\bibitem[\protect\citeauthoryear{Kobayashi}{2002}]{kobD02}
Kobayashi, C., 2002, PhD thesis, Univ. of Tokyo

\bibitem[\protect\citeauthoryear{Kobayashi \& Arimoto}{1999}]{kob99} 
Kobayashi, C., \& Arimoto, N. 1999, \apj, 527, 573

\bibitem[\protect\citeauthoryear{Kobayashi et al.}{1998}]{kob98} 
Kobayashi, C., Tsujimoto, T., Nomoto, K., Hachisu, I, \& Kato, M. 1998, \apj, 503, L155

\bibitem[\protect\citeauthoryear{Kobayashi, Tsujimoto \& Nomoto}{2000}]{kob00} 
Kobayashi, C., Tsujimoto, T., \& Nomoto, K. 2000, \apj, 539, 26

\bibitem[\protect\citeauthoryear{Kodama}{1997}]{kodD97} 
Kodama, T. 1997, Ph.D. Thesis. University of Tokyo

\bibitem[\protect\citeauthoryear{Kodama \& Arimoto}{1997}]{kod97} 
Kodama, T., \& Arimoto, N. 1997, \aap, 320, 41

\bibitem[\protect\citeauthoryear{Kodama et al.}{1998}]{kod98} 
Kodama, T., Arimoto, N., Barger, A.J., \& Arag\'on-Salamanca, A. 1998a, 
\aap, 334,99

\bibitem[\protect\citeauthoryear{Larson}{1974a}]{lar74a}
Larson, R. B. 1974a, MNRAS, 166, 585

\bibitem[\protect\citeauthoryear{Larson}{1974b}]{lar74b}
Larson, R. B. 1974b, MNRAS, 169, 229

\bibitem[\protect\citeauthoryear{Larson}{1975}]{lar75}
Larson, R. B. 1975, MNRAS, 173, 671

\bibitem[\protect\citeauthoryear{Lattanzio et al.}{1985}]{lat85}
Lattanzio, C., Monaghan, J., Pongracic, H., \& Schwarz, P. 1985, \mnras, 215, 125

\bibitem[\protect\citeauthoryear{Leitherer, Robert \& Drissen}{1992}]{lei92}
Leitherer, C., Robert, C., \& Drissen, L. 1992, \apj, 401, 596

\bibitem[\protect\citeauthoryear{Low \& Lynden-Bell}{1976}]{low76}
Low, C., \& Lynden-Bell, D. 1976, \mnras, 176, 367

\bibitem[\protect\citeauthoryear{Lucy}{1977}]{luc77}
Lucy, L. 1977, \aj, 82, 1013

\bibitem[\protect\citeauthoryear{Lynden-Bell}{1967}]{lyn67}
Lynden-Bell, D. 1967, \mnras, 136, 101

\bibitem[\protect\citeauthoryear{Marchant \& Shapiro}{1977}]{mar77}
Marchant, A. B., \& Shapiro, S. L. 1977, \apj, 215, 1

\bibitem[\protect\citeauthoryear{Matteucci}{1996}]{mat96}
Matteucci, F. 1996, Fundamentals of Cosmic Phisics Vol.17, p.283

\bibitem[\protect\citeauthoryear{Menanteau et al.}{1999}]{men99}
Menanteau, F., Ellis, R. S., Abraham, R. G., Barger, A. J., \& Cowie, L. L. 1999, \mnras, 309, 208

\bibitem[\protect\citeauthoryear{Mosconi et al.}{2001}]{mos01}
Mosconi, M. B., Tissera, P. B., Lambas, D. G., \& Cora, S. A., 2001, \mnras, 325, 34

\bibitem[\protect\citeauthoryear{Monaghan}{1992}]{mon92}
Monaghan, J. J. 1992, \araa, 30, 534

\bibitem[\protect\citeauthoryear{Monaghan \& Gingold}{1983}]{mon83}
Monaghan, J. J., \&  Gingold, R. A. 1983, J. Comput. Phys., 52, 374

\bibitem[\protect\citeauthoryear{Monaghan \& Lattanzio}{1985}]{mon85}
Monaghan, J. J., \& Lattanzio, C. 1985, \aap, 149, 135

\bibitem[\protect\citeauthoryear{Nakasato}{2000}]{nakD00}
Nakasato, N. 2000, PhD thesis, Univ. of Tokyo

\bibitem[\protect\citeauthoryear{Nakasato \& Nomoto}{2003}]{nak03}
Nakasato, N., \& Nomoto, K. 2003, \apj, in press (astro-ph/0301404)

\bibitem[\protect\citeauthoryear{Navarro \& Steinmetz}{1997}]{nav97}
Navarro, J. F., \& Steinmetz, M. 1997, \apj, 478, 13

\bibitem[\protect\citeauthoryear{Navarro \& Steinmetz}{2000}]{nav00}
Navarro, J. F., \& Steinmetz, M. 2000, \apj, 538, 477

\bibitem[\protect\citeauthoryear{Navarro \& White}{1993}]{nav93}
Navarro, J. F., \& White, S. D. M. 1993, \mnras, 265, 271

\bibitem[\protect\citeauthoryear{Navarro \& White}{1994}]{nav94}
Navarro, J. F., \& White, S. D. M. 1994, \mnras, 267, 401

\bibitem[\protect\citeauthoryear{Nomoto et al.}{1997a}]{nom97a}
Nomoto, K., Hashimoto, M, Tsujimoto, T, Thielemann, F. -K, Kishimoto, N., Kubo, Y., \&  Nakasato, N. 1997a, Nuclear Physics, A616, 79c

\bibitem[\protect\citeauthoryear{Nomoto et al.}{1997b}]{nom97b}
Nomoto, K., Iwamoto, K., Nakasato, N., Thielemann, F. -K, Brachwitz, F., Tsujimoto, T., Kubo, Y., \&  Kishimoto, N. 1997b, Nuclear Physics, A621, 467c

\bibitem[\protect\citeauthoryear{Nomoto et al.}{1984}]{nom84}
Nomoto, K., Thielemann, F. -K, \& Yokoi, K. 1984, \apj, 286, 644

\bibitem[\protect\citeauthoryear{Raiteri, Villata \& Navarro}{1996}]{rai96}
Raiteri, C. M., Villata, M., \& Navarro, J. F., \aap, 315, 105

\bibitem[\protect\citeauthoryear{Renzini}{2002}]{ren02}
Renzini, A. 2002, in ASP Conference Series, 253, Chemical Enrichment of Intracluster and Intergalactic Medium, eds., R. Fusco-Femiano \& F. Matteucci, p.331

\bibitem[\protect\citeauthoryear{Salpeter}{1955}]{sal55}
Salpeter, E. E. 1955, \apj, 121, 161

\bibitem[\protect\citeauthoryear{Schade, Barrientos \& Lopez-Cruz}{1997}]{sch97} 
Schade, D., Barrientos, L. F., \& Lopez-Cruz, O. 1997, \apj, 477, L17

\bibitem[\protect\citeauthoryear{Schade et al.}{1999}]{sch99}
Schade, D., et al. 1999, \apj, 525, 31

\bibitem[\protect\citeauthoryear{Schmidt}{1959}]{sch59}
Schmidt, M. 1959, \apj, 129, 243

\bibitem[\protect\citeauthoryear{Schweizer \& Seitzer}{1992}]{sch92} 
Schweizer, F., \& Seitzer, P. 1992, \aj, 104, 1039

\bibitem[\protect\citeauthoryear{Schweizer et al.}{1990}]{sch90} 
Schweizer, F., Seitzer, P., Faber, S. M., Burstein, D., Ore, C. M. D., \& Gonzalez, J. J. 1990, \apj, 364, L33

\bibitem[\protect\citeauthoryear{Silk}{1977}]{sil77}
Silk, J. 1977, \apj, 214, 718

\bibitem[\protect\citeauthoryear{Silva \& Bothun}{1998}]{sil98}
Silva, D. R., \& Bothun, G. D. 1998, \aj, 116, 85

\bibitem[\protect\citeauthoryear{Springel \& Hernquist}{2003}]{spr03} 
Springel, V. \& Hernquist, L. 2003, \mnras, 339, 289

\bibitem[\protect\citeauthoryear{Stanford, Eisenhardt \& Dickinson}{1998}]{sta98} 
Stanford, S. A., Eisenhardt, P. R. M., \& Dickinson, M. 1998, \apj, 492, 461

\bibitem[\protect\citeauthoryear{Steinmetz}{1996}]{ste96}
Steinmetz, M. 1996, \mnras, 278, 1005

\bibitem[\protect\citeauthoryear{Steinmetz \& M\"uller}{1994}]{ste94}
Steinmetz, M., \& M\"uller, E. 1994, \aap, 281, L97

\bibitem[\protect\citeauthoryear{Steinmetz \& Navarro}{2002}]{ste02}
Steinmetz, M., \& Navarro, J. F. 2002, New Astronomy, 7, 155

\bibitem[\protect\citeauthoryear{Steinmetz \& White}{1997}]{ste97}
Steinmetz, M., \& White, S. D. M. 1997, \mnras, 288, 545

\bibitem[\protect\citeauthoryear{Sugimoto et al.}{1990}]{sug90}
Sugimoto, D., Chikada, Y., Makino, J., Ito, T., Ebisuzaki, T., 
\& Umemura, M. 1990, Nature, 345, 33

\bibitem[\protect\citeauthoryear{Sutherland \& Dopita}{1993}]{sut93}
Sutherland, R. S., \& Dopita, M. A. 1993, \apjs, 88, 235

\bibitem[\protect\citeauthoryear{Thomas \& Couchman}{1992}]{tho92}
Thomas, P. A., \& Couchman, H. M. P. 1992, \mnras, 257, 11

\bibitem[\protect\citeauthoryear{Tinsley}{1980}]{tin80}
Tinsley, B. M. 1980, Fundamentals of Cosmic Phisics Vol.5, p.287

\bibitem[\protect\citeauthoryear{Toomre \& Toomre}{1972}]{too72}
Toomre, A., \& Tommre, J. 1972, ApJ, 178, 623

\bibitem[\protect\citeauthoryear{Toomre}{1977}]{too77}
Toomre, A. 1977, in The Evolution of Galaxies and Stellar Populations, ed. B. M. Tinsley and R. B. Larson (New Haven: Yale University Observatory), p.401

\bibitem[\protect\citeauthoryear{Tripicco \& Bell}{1995}]{tri95} 
Tripicco, M. J., \& Bell, R. A. 1995, \aj, 110, 3035

\bibitem[\protect\citeauthoryear{van Dokkum \& Franx}{1996}]{van96}
van Dokkum, P. G., \& Franx, M. 1996, \mnras, 281, 985

\bibitem[\protect\citeauthoryear{van Dokkum \& Franx}{2001}]{van01}
van Dokkum, P. G., \& Franx, M. 2001, \apj, 553, 90

\bibitem[\protect\citeauthoryear{Umemura et al.}{1993}]{ume93}
Umemura, M., Fukushige, T., Makino, J., Ebisuzaki, T., Sugimoto, D., 
Turner, E., \& Loeb, A. 1993, \pasj, 45, 311

\bibitem[\protect\citeauthoryear{van Albada}{1982}]{vanalb82}
van Albada, T. S., 1982, \mnras, 201, 939

\bibitem[\protect\citeauthoryear{Warren et al.}{1992}]{war92}
Warren, M. S., Quinn, P. J., Salmon, J. K., \& Zurek, W. H. 1992, \apj, 399, 405

\bibitem[\protect\citeauthoryear{White}{1980}]{whi80}
White, S. D. M. 1980, \mnras, 191, 1p

\bibitem[\protect\citeauthoryear{White}{1978}]{whi78}
White, S. D. M. 1978, \mnras, 184, 185

\bibitem[\protect\citeauthoryear{Woosley \& Weaver}{1995}]{woo95} 
Woosley, S. E., \& Weaver, T. A. 1995, \apjs, 101, 181

\bibitem[\protect\citeauthoryear{Zel'dovich}{1970}]{zel70}
Zel'dovich, Ya. B. 1970, \aap, 5, 84

\bibitem[\protect\citeauthoryear{Zepf}{1997}]{zep97}
Zepf, S. E. 1997, Nature, 390, 377

\end{thebibliography}
\end{document}